\documentclass[aps,prb,floatfix,twocolumn,superscriptaddress,amssymb,amsmath]{revtex4-1}
\usepackage{graphicx}
\usepackage{dcolumn}% Align table columns on decimal point
\usepackage{bm}% bold math
\usepackage{amsmath}
\usepackage{color}
\usepackage{mathrsfs}
\usepackage{float}
\usepackage{subfigure}
\usepackage{dsfont}
\usepackage{txfonts}
\usepackage{wasysym}
\usepackage{ifsym}

\usepackage{multirow}
\makeatletter

\newcommand{\Rmnum}[1]{\expandafter\@slowromancap\romannumeral #1@}
\makeatother
\def \beq {\begin{equation}}
\def \edq {\end{equation}}
\def \ba {\begin{eqnarray}}
\def \ea {\end{eqnarray}}
\def \bes {\begin{subequations}}
\def \eds {\end{subequations}}
\def \beqn {\begin{equation*}}
\def \edqn {\end{equation*}}

\begin{document}

\title{Yu-Shiba-Rusinov multiplets and clusters of multiorbital adatoms  in superconducting substrates: Subgap Green's function approach}

\author{Liliana Arrachea}
\affiliation{International Center for Advanced Studies, Escuela de Ciencia y Tecnolog\'{\i}a and ICIFI, UNSAM, Campus Miguelete, 25 de Mayo y Francia, 1650 Buenos Aires, Argentina}

\begin{abstract}
We discuss all the characteristics of Yu-Shiba-Rusinov states for clusters of  impurities with classical magnetic moments in a superconducting substrate with $s$-wave symmetry. We consider the effect of the multiorbital structure of the impurities and the effect of the crystal field splitting. We solve the problem exactly and calculate the {\em subgap Green's function}, which has poles at the energies of the Shiba states and defines the local density of states associated to their wave functions.  For the case of impurities sufficiently separated, we derive an effective Hamiltonian to describe the hybridization mediated by the substrate. We analyze the main features of the spectrum and the 
spectral density of the subgap excitations for impurities in dimer configurations with different relative orientations of the magnetic moments. We also illustrate how the same formalism applies for the solution of a trimer with frustration in the orientation of the magnetic moments. 
\end{abstract}

\date{\today}

\maketitle

%%%%%%%%%%%%%%%%%%%%%%%%%%%%%%%%%%%%%%%%%%%%%%%%%%%
\section{Introduction}
The study of magnetic impurities in superconducting hosts is receiving a significant attention for some years now in the context of
conventional and non-conventional materials. \cite{balatsky} 
A single classical impurity coupled through a magnetic exchange interaction  $J$ to a superconductor leads to the formation of  Yu-Shiba-Rusinov states (YSR),
which are spatially localized around the impurity and have energies in the superconducting gap of the bulk, $\Delta$. \cite{ysr1,ysr2,ysr3} 
 For a singlet-type superconductor, they
inherit the symmetries of the  environment, 
 appearing in pairs of quasiparticles with oposite energies. They are spin-polarized parallel or antiparallel to the classical magnetic moment ${\bf S}$. 
 As the exchange interaction $J$ increases, the energies of the YSR states evolve from $\pm \Delta$ to zero until the exchange interaction overcomes a critical value $J^c$, where the two states cross at zero energy. This defines a quantum phase transition, where
 the parity of the  ground state changes, as has been discussed in many places in the literature. \cite{balatsky,hein}.

 \begin{figure}[t]\begin{center}
  \includegraphics[width=\columnwidth,height=7cm]{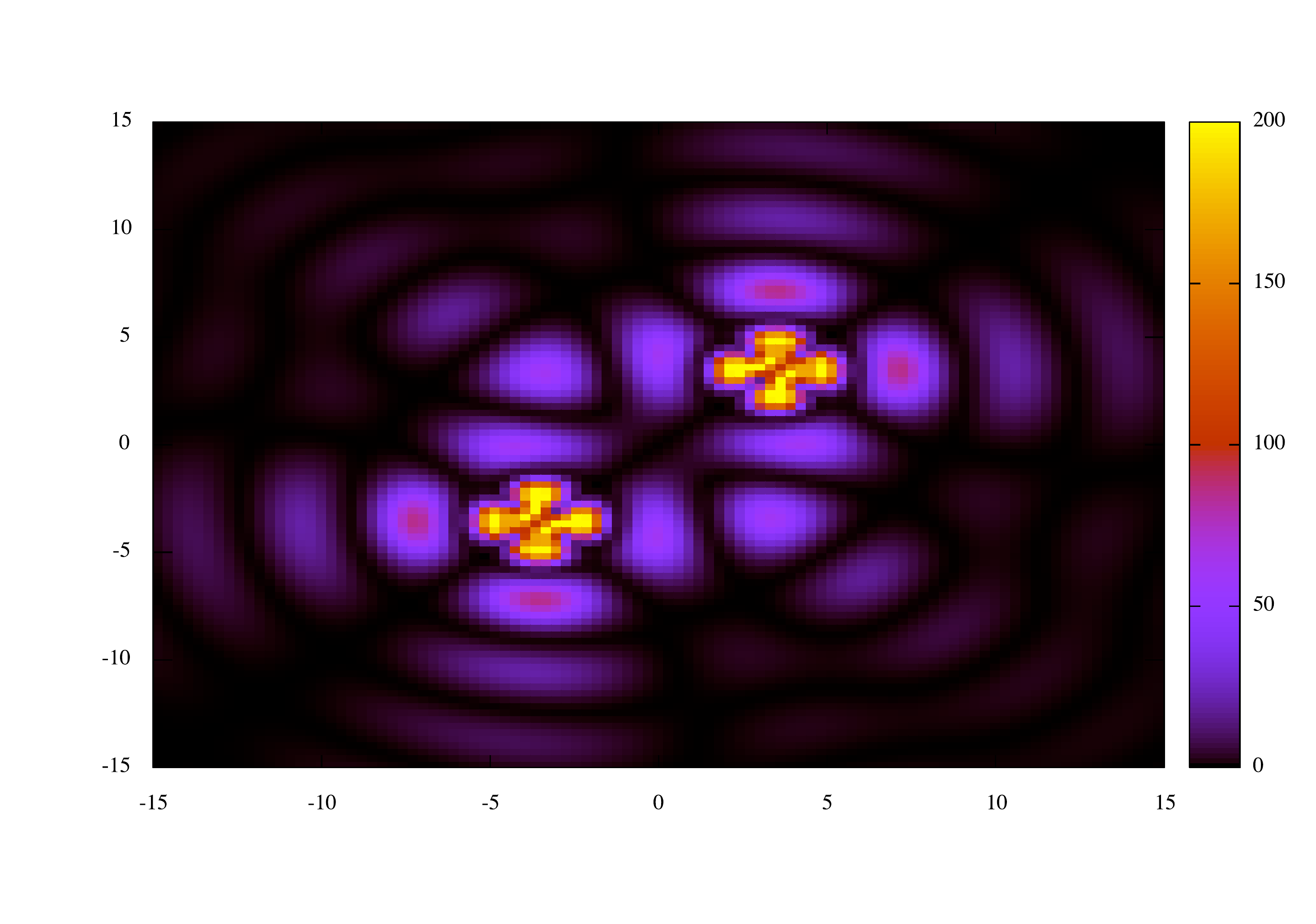}
    \caption{Spectral density of the YSR state with bonding struture of a dimer with parallel classical magnetic moments. The orbital channel is $d_{x^2-y^2}$.}
 \label{figilu}
\end{center}\end{figure}
The quantum phase transition  of the classical magnetic impurity bears a  resemblance to the Kondo effect of quantum magnetic impurities in superconducting hosts. For impurities with spin $S=1/2$, a Kondo temperature $T_K$ is defined such that for $k_B T_K \ll  \Delta$ the impurity behaves as a free spin embedded in a superconductor, while
 for $k_B T_K \gg  \Delta$ the impurity forms a singlet with bound quasiparticle states. In the former case, singlet combinations of the impurity with quasiparticle states form subgap excitations. In the latter case, there are also subgap excitations, resulting  from the broken Kondo singlet. These features are observed in calculations with numerical renormalization group of the Anderson impurity model in a superconducting substrate, \cite{zitko1,zitko2} as well as experimentally in quantum dots, which behave like magnetic impurities. \cite{deacon,agua}

Nadj-Perge and coworkers proposed the construction of artificial structures of magnetic adatoms to realize topological superconductivity, \cite{nadj1} which motivated a pletora of recent studies. \cite{nadj2,pletorap1,pletora1,pletora2,PGvO,nico,alf1,felix}  Magnetic adatoms  usually have
$d$ or $f$ external orbitals. Hence, the impurity may have a more complex structure than a classical magnetic moment coupled to the superconducting substrate. In Ref. \onlinecite{moca} the Shiba spectrum of a Mn impurity with 3d active electrons
 in a two-band superconductor was studied taking into account the atomic orbital degrees of freedom, in addition to the spin.  More recently,  features related to the orbital structure have been observed  with scanning tunneling microscope (STM) experiments. \cite{yang,dj} 
 In these systems,
 the net magnetic moment of the adatoms is originated in a dominant internal Hund rule, while the coupling of the orbital degrees of freedom define multiple channels in the Kondo problem. 
 In normal metal (non-superconducting) substrates, these effects 
 have been studied in the context of the two-orbital Anderson impurity model, corresponding to $S=1$ quantum impurities theoretically, \cite{mk1,mk2,mk3} as well as experimentally.\cite{mk4,mk5} The additional ingredient that the multi-orbital structure brings about is the possibility of underscreening, in addition to
the  full screening mechanism taking place in the single-orbital case.

  The search for topological 
 phases in structures of YSR states demands the construction of complex arrays containing several adatoms. 
 The simplest of these arrays is a dimer containing two adatoms.
Results for  a dimer of classical impurities in a s-wave superconductor have  been presented in Refs. \onlinecite{flatte,meng1,meng2} and  the phase diagram of a dimer of quantum  ${\bf S}=1/2$ impurities was studied in Ref. \onlinecite{dimer}, while trimers of ${\bf S}=1/2$ impurities have been studied in normal metal \cite{artri} and superconducting\cite{koerber} substrates. All these works focus on adatoms with a single active orbital.
The YSR states localized at the impurities in clusters of multiorbital adatoms hybridize through the substrate and many scenarios and phase transitions may take place. 
In a recent work the transport properties of two impurities realized in a double quantum-dot structure embbeded in a superconducting Josephson junction were analized. \cite{alf}
Experimental research with STM spectroscopy on dimers of Cr and Mn atoms  on superconducting substrates has been recently reported in Refs. \onlinecite{nacho,katha}. Even more recently, 
dimers and trimers of Gd atoms  in superconductors have been also studied with STM techniques in Ref. \onlinecite{ding}.
Maps like the one shown in Fig. \ref{figilu} are recorded in these experiments, which 
 reveal details on the multiorbital structure of the adatoms and the hybridization between the impurities mediated by the substrate.

The aim of this
 work is to present a systematic theoretical framework to analyze  the different scenarios expected to take place when clusters of adatoms with several active orbitals are placed in superconducting substrates, considering the net spin of the atoms as classical magnetic moments. We base this description on the definition of an effective Green's function to describe the bound subgap states, which can be exactly evaluated. In addition, we derive 
 an effective Hamiltonian 
 to calculate the subgap excitations for the case where the impurities are sufficiently diluted. In both cases, not only the spectrum but also the density of states of the YSR states can be calculated and analyzed, taking into account
 the relative orientation of the magnetic moments, as well as the multiorbital structure of the adatoms and the effect of the crystal field generated by the environment. 
 We illustrate the formalism mainly focusing on dimers with d orbitals and a substrate with a constant density of states, but we also revisit the single multiorbital impurity and a trimer configuration. The general formalism applies to any type of orbitals and can be extended to describe other type of substrates.

 The paper is organized as follows.  In section \ref{sec:model} we present details of the derivation of the model describing the hybridization of the multiple orbitals of the impurity with the substrate and we briefly review the derivation of the low energy Hamiltonian for the impurity originally presented in Ref. \onlinecite{blandin}. We also discuss with intuitive arguments 
 how the picture of the quantum phase transition taking place as a function of $J$ generalizes when the impurity has several active orbitals and when instead of a single impurity, there are two impurities forming a dimer. 
 This model can be exactly solved with the methods presented in sections \ref{sec:formal}, where  we present the derivation of an effective Hamiltonian for the case of diluted impurities.
Results are presented in section \ref{sec:res} and section \ref{sec:sum-con} is devoted to summary and conclusions. The appendices contain some technical details.

\section{Theoretical description}\label{sec:model}
\subsection{Model}
Typically, magnetic adatoms have up to  $2l+1$ active orbitals of  angular momentum $l$ that hybridize with the substrate, and  a strong intra-atomic Hund rule generating a magnetic moment with total spin $S=(2l+1)/2$. The low-energy
model for such an impurity  in metallic substrates was originally derived in a seminal paper by Nozi\'eres and Blandin, \cite{blandin} who
 considered the most general Anderson impurity model in real metals.  Taking into account the orbital structure of the electrons of the impurity that hybridize to the substrate, and the  crystal field splitting, they derived the low-energy Hamiltonian where the impurity is represented by a quantum spin in a conducting environment. The latter defines the model to investigate the Kondo effect in realistic scenarios. In what follows, we
adapt such derivation to the case of a superconducting substrate. In a second step, we treat the net spin of the impurity as a classical magnetic moment. 
 For simplicity, we will mainly focus on a BCS model for the  superconducting  host expressed in a basis of plane waves, assuming a pairing interaction with s-wave symmetry.
 The same steps of our reasoning can be adapted to the
case of a substrate of Bloch electrons and to other symmetries of the superconducting order parameter.

%\subsection{Single impurity with several atomic orbitals}
It is instructive to derive the Hamiltonian for the multiorbital impurity in the superconducting host following the steps of Ref. \onlinecite{tsvelic}. The starting point is the definition of a convenient basis
to adequately represent the delocalized degrees of freedom of the substrate and the localized ones at the impurity. The natural basis for the substrate 
is that of plane waves.  In 3D,  plane waves can be expanded in spherical waves as follows,
\begin{equation} \label{plane}
\psi_{l,m}({\bf k},{\bf r}) =\frac{4 \pi  (i)^l}{\sqrt{\cal V}}  j_l(kr) Y^m_l(\hat{n}_k) Y_l^m(\hat{n}_r),
\end{equation}
where ${\cal V}$ is the volume of the system and $j_l(kr)$ is the spherical Bessel function of order $l$. On the other hand, for the impurity it is natural to  consider a basis 
constructed with atomic wave functions with well defined angular momentum $l$,  $\phi^{imp}_{l,m}({\bf r}) = Y_{l,m}(\hat{n}_r) \phi(r) $, being   $Y_{l,m}(\hat{n}_r)$ 
the spherical harmonics with $m=-l, \ldots, l$. 

The  field operator for an electron with spin $\sigma$  can be expanded as follows,
\begin{equation} \label{p3d}
\Psi_{\sigma}({\bf r}) = {\sum_{{\bf k},l,m}}^{\prime} \psi_{l,m}({\bf k},{\bf r}) {\bf c}_{\bf k,\sigma} + \sum_{l,m} \phi^{imp}_{l,m}({\bf r}) {\bf d}_{l,m,\sigma}.
\end{equation}
The $\sum^{\prime}_{\bf k}$ denotes summation over ${\bf k}$-vectors with $k < r_B^{-1}$, being $r_B$ the Bohr radius, which prevents problems  related to the fact that 
the two basis used in the expansion above are not orthogonal. \cite{tsvelic} 

The Hamiltonian  for the impurity in the superconducting host expressed in terms of these field operators reads
\begin{eqnarray} \label{ham}
H &=&\int d^3 r \left\{ \sum_{\sigma}  \Psi_{\sigma}^{\dagger}({\bf r})  \left[ {\cal H}^0_{\sigma,{\rm sup}}+{\cal H}^0_{\sigma, {\rm imp}} \right] \Psi_{\sigma}({\bf r}) \right. \nonumber\\
& & \left. \;\;\;\;\;\;+ \Delta \Psi_{\uparrow}^{\dagger}({\bf r})\Psi_{\downarrow}^{\dagger}({\bf r}) + H. c. \right\}+H^{\rm Coul}_{\rm imp},
\end{eqnarray}
where $H^0_{\rm sup}$ corresponds to the kinetic term of the Hamiltonian for the 
superconducting substrate and $H^0_{imp}$ is the non-interacting  Hamiltonian of the impurity, the first term of the second line is the local usual s-wave BCS-pairing  Hamiltonian 
and $H^{\rm Coul}_{\rm imp}$ is the Coulomb interaction for the electrons in the impurity. 

Substituting  the expansion of Eq. (\ref{p3d}) in the latter Hamiltonian we get an expression in terms of the creation and annihilation operators, which reads
\begin{eqnarray} \label{hamin}
H &= & \sum_{{\bf k}} \left[ \sum_{\sigma} \xi_{\bf k} c^{\dagger}_{{\bf k},\sigma} c_{{\bf k}, \sigma} + \Delta c^{\dagger}_{{\bf k},\uparrow} c^{\dagger}_{- {\bf k},\downarrow} + H.c.\right]+ H_{\rm hyb}
 \nonumber \\
&  &+ H_{\rm imp} \left[ d^{\dagger}_{l,m,\sigma}, d_{l,m,\sigma} \right].
\end{eqnarray}
Here $H_{\rm imp}$ collects all the terms depending on the operators $d^{\dagger}_{l,m,\sigma}, \; d_{l,m,\sigma} $ describing the orbital and spin degrees of freedom of the impurity.
The term $H_{\rm hyb}$ describes the hybridization terms between the degrees of freedom of the substrate and the ones in the impurity. 
The corresponding matrix elements are obtained by computing
\begin{eqnarray}\label{wiso}
& &  \int dr r^2 d\Omega_r \; \psi^*_{l^{\prime},m^{\prime}}({\bf k},{\bf r}) \left[ {\cal H}^0_{\rm sup}+{\cal H}^0_{\rm imp} \right] \phi^{imp}_{l,m}({\bf r})  \nonumber \\
& & \simeq
W_{k,l,m} Y^m_l(\hat{n}_k)\delta_{l,l^{\prime}} \delta_{m,m^{\prime}}.
\end{eqnarray}
We have used the normalization of the spherical harmonics and we have introduced the parameter
\begin{equation}
W_{k,l,m} =   \frac{4 \pi (i)^l}{\sqrt{\cal V}}   \int dr r^2 j_l(kr) \left[ {\cal H}^0_{\rm sup}+{\cal H}^0_{\rm imp} \right]  \phi_{l,m}(r),
\end{equation}
which defines the hybridization amplitude between the atomic orbital with angular momentum $l$ and the substrate. Thus, the hybridization Hamiltonian reads
\begin{equation}
H_{\rm hyb}=  \sum_{{\bf k}, l,m,\sigma} W_{k,l,m} Y^m_l(\hat{n}_k) c^{\dagger}_{{\bf k},\sigma} d_{l,m,\sigma} + H. c.
\end{equation}
The latter  expression  is the starting point of Ref. \onlinecite{blandin}, where it is argued that 
that the hybridization parameter is approximately the same for all $k$, i.e., $W_{k,l,m}\sim W_{l,m}$.
 
 When the  impurity is embedded in a lattice, rotational symmetry is broken. Hence,  the degenerate levels $m$,  are split by the electric 
potential of the neighboring atoms and $m$ is no longer a good quantum number. This is, precisely, the effect of the crystal field splitting. In this context, the internal atomic levels are more appropriately described by a label which depends on the irreducible representation of the point group of symmetry at the impurity. We will focus on the case of impurities with  $d$ electrons ($l=2$), in which case, it is usual to introduce the labeling of the cubic harmonics, $\mu= d_{x^2-y^2},d_{ z^2}, d_{xy}, d_{xz}, d_{yz}$,
instead of $m$. In a system with crystal field splitting, the hybridization parameter is not expected to be the same for all the atomic channels and it depends on the representation label characterizing the  different channels,
$W_{\mu}$. Hereafter, we assume that the crystal field is relevant, hence we label with the index $\mu$. 

Following again Ref. \onlinecite{blandin}, it is possible to 
eliminate
the high-energy states of the impurity by recourse to a Schrieffer-Wolff transformation. Such a procedure consists in integrating out the (high-energy) states associated to the charge fluctuations in the impurity. 
In the case where all the  relevant orbitals labeled by $\mu$ are filled, half-filled or empty,
this
leads to a Kondo Hamiltonian, where the many-body states of the impurity are projected only on those describing its total spin. If 
 there is a strong Hund rule in the impurity, its total spin is large and it is usual to simplify the low-energy Hamiltonian by representing the spin impurity ${\bf S}$ by a classical magnetic moment.
Typically  Mn$^{+2}$, V$^{+2}$ or Ni$^{+2}$ in cubic symmetry satisfy these conditions. 
All these considerations lead to the low-energy Hamiltonian  $H_{\rm low}=H_{\rm sup}+ H_{\rm int}$ which
replaces $H$ of Eq. (\ref{hamin}). 
 At this point it is also convenient to introduce Nambu notation for the fermionic operators of the substrate ${\bf c}_{\bf k} = \left(c_{\bf k,\uparrow}, c_{\bf k,\downarrow}, c^{\dagger}_{-{\bf k},\downarrow},
 - c^{\dagger}_{-{\bf k},\uparrow}\right)^T$. The different terms of $H_{\rm low}$ read
\begin{eqnarray}\label{hams-w}
H_{\rm sup} &=& \sum_ {\bf k} {\bf c}^{\dagger}_{\bf k} \left[ \xi_{\bf k} \tau_z + \Delta \tau_x \right]  \;{\bf c}_{\bf k}, \nonumber \\
H_{\rm int} &=& 
\sum_{{\bf k}, {\bf k}^{\prime}}  
{\bf c}^{\dagger}_{\bf k} \; V({\bf k},{\bf k}^{\prime}) \;
% \left[\; U_{\mu} \tau_z + J_{\mu}  \;  \;  \boldsymbol{\sigma} \cdot {\bf S}\; \right]f^*_{\mu}({\bf k})f_{\mu}({\bf k}^{\prime})
 {\bf c}_{{\bf k}^{\prime}}.
\end{eqnarray}
with
\begin{eqnarray}\label{vkk}
V({\bf k},{\bf k}^{\prime}) &=& \sum_{\mu} f^*_{\mu}({\bf k}) V_{\mu}  f_{\mu}({\bf k}^{\prime})\nonumber \\
V_{\mu} &=&  \left[\; U_{\mu} \tau_z + J_{\mu}  \;  \;  \boldsymbol{\sigma} \cdot {\bf S}\; \right].
\end {eqnarray}
Here $\tau_{x,y,z}$ are Pauli matrices acting in he particle-hole degrees of freedom of the Nambu spinor, while $ \boldsymbol{\sigma}=\left(\sigma_x,\sigma_y,\sigma_z\right)$ are Pauli matrices
acting on the spin degrees of freedom. 
The first term of $H_{\rm int}$
is the potential scattering and  the second one is the exchange interaction between the total spin of the impurity,
 ${\bf S}$ (described by a classical vector), and the spin of the electrons in the substrate. 
 The parameters entering the interaction depend on the nature of the substrate and the energy for the impurity charge fluctuations $\Delta E$ as follows
  \begin{equation}
J_{\mu} ,U_{\mu} =     \frac{ W_\mu^2 }{\Delta E}.
\end{equation}
The coefficients $f_{\mu}$ in Eq. (\ref{vkk}) are in general combinations of the spherical harmonic functions. 
For the case of $d$ orbitals and a plane-wave substrate, they are the cubic harmonics
\begin{eqnarray}\label{repre}
&  & f_{z^2}=Y_2^0,\;\;\; f_{x^2-y^2}=\frac{1}{\sqrt{2}}\left( Y_2^{-2}+Y_2^2 \right),\;\;\;f_{xy} =-\frac{i}{\sqrt{2}} \left( Y_2^{-2}- Y_2^2 \right), \nonumber \\
& & f_{xz} =\frac{1}{\sqrt{2}} \left(Y_2^{-1}-Y_2^1 \right),\;\;\;
f_{yz} =-\frac{i}{\sqrt{2}} \left(Y_2^{-1}+Y_2^1 \right).
\end{eqnarray}

%\subsection{Several impurities}
We now extend the previous reasoning to the case of  $N_I$ impurities at positions ${\bf r}_j, \; j=1, \ldots, N_I$, which we assume to be sufficiently separated so that we can neglect any direct hybridization between them. 
%We will see that an effective interaction is induced due to the hybridization of the individual impurities with the substrate. 
The basis -- equivalent to Eq. (\ref{p3d}) -- to express the  Hamiltonian in terms of creation and annihilation operators is now
\begin{equation} \label{p3d-2}
\Psi_{\sigma}({\bf r}) = \sum^{\prime}_{{\bf k},\sigma} \psi_{\mu}({\bf k},{\bf r}) c_{{\bf k},\sigma} + \sum_{j=1}^{N_I} \sum_{l,m} \phi^{\rm imp}_{j, l,m }({\bf r}) d_{j, l,m, \sigma},
\end{equation}
The basis of functions localized at the impurities are
%\begin{equation}
$\phi^{\rm imp}_{j, l,m}({\bf r} )= Y_{l,m}(\hat{n}_{{\bf r}- {\bf r}_j}) \; \phi(|{\bf r}- {\bf r}_j|)$,
%\end{equation}
where $\hat{n}_{{\bf r}- {\bf r}_j}$ are the angular coordinates of the vector ${\bf r}- {\bf r}_j$ and $Y_{l,m}(\hat{\mu}_{{\bf r}- {\bf r}_j})$ is a spherical harmonic. For a system with a strong crystal field 
splitting, it is a combination of spherical harmonics corresponding to the relevant irreducible representation of the point symmetry group. $\phi(|{\bf r}- {\bf r}_j|)$ is a function strongly localized at the impurity.

After following the same steps as before, we derive the  low-energy Hamiltonian analogous to Eq. (\ref{hams-w}). The corresponding interaction matrix reads
\begin{eqnarray}\label{vkksev}
V({\bf k},{\bf k}^{\prime}) &=& \sum_{j=1}^{N_I} \sum_{\mu} {f^j_{\mu}}({\bf k})^* V^j_{\mu}  f^j_{\mu}({\bf k}^{\prime})\nonumber \\
V^j_{\mu} &=&  \left[\; U^{j}_{\mu} \tau_z + J^{j}_{\mu}  \;  \;  \boldsymbol{\sigma} \cdot {\bf S}_j\; \right].
\end {eqnarray}
being ${\bf S}_j $ the classical vector representing the magnetic moment of the $j$-th impurity while
\begin{equation}
U^j_{\mu}, J^j_{\mu}  =     W_j^2 /\Delta E, \;\;\;\;\;\;\;\;\;\;\;\; f^j_{\mu}({\bf k}) \equiv e^{-i {\bf k} \cdot {\bf r}_j }Y_{\mu} (\hat{n}_{k}),
\end{equation}
with all the parameters having the same meaning as in the case of the simple impurity. Notice that, as before, we have substituted the indices $l,m$ by the representation index $\mu$, assuming the
effect of the crystal field splitting.

\section{Formalism}\label{sec:formal}

\subsection{Green's function and T-Matrix} \label{sec:green-t}
We introduce the Nambu notation for the field operators of the electrons in the substrate, $\Psi({\bf r})= \left( \Psi_{\uparrow}({\bf r}), \Psi_{\downarrow}({\bf r}), \Psi^{\dagger}_{\downarrow}({\bf r}), - \Psi^{\dagger}_{\uparrow}({\bf r}) \right)$.
The corresponding retarded Green's function in terms of the T-matrix can be expressed as follows
\begin{eqnarray}\label{greenr}
G({\bf r}, {\bf r}, \omega) &=&  G^0(0,\omega) +  \\
& & \sum_{\mu,\mu^{\prime}} \sum_{j,j^{\prime}} \gamma_{\mu}({\bf r} -  {\bf r}_j, \omega)^* \; T^{j,j^{\prime}}_{\mu,\mu^{\prime}}(\omega) \; \gamma_{\mu^{\prime}}({\bf r}_{j^{\prime}}- {\bf r},\omega),
\nonumber
\end{eqnarray}
where $ G^0(0,\omega)$ is the local Green's function of the substrate without impurities.  The matrices $\gamma_{\mu}({\bf r},\omega)$ are 
\begin{equation}\label{gamdef}
\gamma_{\mu}({\bf r}, \omega) = \sum_{\bf k} e^{i {\bf k \cdot r}} G^0({\bf k}, \omega) f_{\mu}({\bf k})^*,
\end{equation}
and the $T$-matrix  is determined from the equation 
\begin{equation} \label{tmjj}
T^{j,j^{\prime}}_{\mu,\mu^{\prime}} (\omega) = \delta_{j,j^{\prime}} \delta_{\mu,\mu^{\prime}} V_{\mu}^j +\sum_{j^{\prime \prime},\mu^{\prime \prime}} T_{\mu,\mu^{\prime \prime}}^{j,j^{\prime \prime}}(\omega) F_{\mu^{\prime \prime},\mu^{\prime}}^{j^{\prime \prime}, j^{\prime}}(\omega) V_{\mu^{\prime}}^{j^{\prime}}, 
\end{equation}
with
\begin{equation} \label{fjjp}
F_{\mu,\mu^{\prime}}^{j, j^{\prime}}(\omega) = \sum_{\bf k} f^j_{\mu}({\bf k})^*  G^0({\bf k}, \omega) f^{j^{\prime}}_{\mu^{\prime}} ({\bf k}). 
\end{equation}
The eigenenergies of the Shiba states are determined from the energies of the poles of the $T$-matrix within the gap. 
It is interesting to notice that the $T$-matrix of Eq. (\ref{tmjj}) 
couples different impurities and different channels, in spite of the fact that  the  bare interaction $V$ is diagonal in these indices. This represents an effective interaction mediated by the substrate that is not present in the original model and plays 
a key role in the nature of the hybridization of the Shiba states. 
In the case of a {\em single impurity}, due to the orthogonality of the functions $f^j_{\mu}({\bf k})$, the function $F$ reduces to 
\begin{equation}\label{def-f}
F(\omega)=\sum_{\bf k}  G^0({\bf k}, \omega) 
\end{equation}
 while the $T$-matrix is diagonal in the orbital index and reads 
\begin{equation}\label{tmdef}
T_{\mu}(\omega)= V_{\mu} \left[ 1- F(\omega) V_{\mu} \right]^{-1}.
\end{equation}

\subsection{Subgap Green's function and  Shiba states} \label{sec:green-b}
The aim of the present section is to define an
effective Green's function $\hat{\cal G} (\omega)$ to represent the Shiba states. We define it as follows,
\begin{equation}\label{tg}
\hat{T}(\omega) = \hat{V} \; \hat{\cal G} (\omega) \; \hat{V}\;\;\;\;\;\;\;\;\; |\omega|<\Delta,
\end{equation}
where we $\hat{T}(\omega)$ and $\hat{V}$ are matrices in the impurity, orbital and Nambu indices. 
In this way,  this Green's function has poles at the Shiba energies and the corresponding quasiparticle weight defines the  contribution of the Shiba states to the local density of states. 

\subsubsection{Single impurity}
We consider the $T$-matrix defined in Eq. (\ref{tmdef}). 
The eigenenergies of the Shiba states correspond to the poles of this matrix  within the range  $|\omega| < \Delta$, and 
are determined from the condition $\mbox{Det}[1-F_{\mu}(\omega) V_{\mu}]=0$.

We are now interested in constructing an effective  Green's function ${\cal G}_{\mu}(\omega)$, such that its  poles in the range of energies with $|\omega|<\Delta$ coincide with those of the $T$-matrix. 
We define an {\em auxiliary matrix} such that its zeroes coincide with the zeroes of $ {\cal G}^{-1}_{\mu}(\omega)$, hence with the poles of ${\cal G}_{\mu}(\omega)$,
 \begin{equation}\label{lambda}
 %{\cal G}_{\mu}^{-1}(\omega)
 \Lambda_{\mu}(\omega)= V_{\mu}-V_{\mu}F_{\mu}(\omega) V_{\mu}.
 \end{equation}
  In order to find the zeroes of this matrix, it is convenient to diagonalize it for each $\omega$ and express it in terms of the corresponding 
  eigenvalues $\epsilon_m$ and eigenstates $|\Phi_m \rangle$ as follows,
  \begin{equation}\label{esp-shiba}
\Lambda_{\mu}(\omega) = \sum_m  \epsilon_m(\omega) | \Phi_m(\omega) \rangle \langle \Phi_m (\omega) |.
\end{equation}
The Shiba state $| {\mu, \rm s} \rangle$ with energy $E_{\mu, \rm s}$ corresponds to the condition that one of these eigenstates has vanishing eigenvalue  $\epsilon_m(E_{\mu, \rm s})=0$.
Due to the Nambu structure, they have opposite energies, $E_{\mu, +}= - E_{\mu, -}$, since the two states are related by a charge-conjugation and time reversal (${\rm CT}$)  transformation. 
The latter reads $| \mu, - \rangle ={\rm CT} |\mu, +  \rangle$ with ${\rm C}=-i \tau_y$ and ${\rm T}=i \sigma_y K$, being $K$ is the complex conjugation operation.

Since we are looking for a Green's function with poles at the energies $E_{\mu, \rm s}$, its inverse must satisfy
\begin{eqnarray} \label{shiba}
 {\cal G}^{-1}_{\mu}(E_{\mu, \rm s}) | {\mu, \rm s} \rangle 
& = & \left[ V_{\mu}-V_{\mu}F_{\mu}(E_{\mu, \rm s}) V_{\mu} \right]  | {\mu, \rm s} \rangle =0.
\end{eqnarray}
%Thus, we can easily verify  for this energy we have $\mbox{Det}[1-F_{\mu}(E_{\mu, \rm s})]=0$, implying that it is a pole of the $T$-matrix. 
In order to get an expression for ${\cal G}^{-1}_{\mu} (\omega)$ in a neighborhood of  the Shiba energy,
 $\omega \simeq E_{\mu, \rm s}$, we perform the following expansion
 \begin{equation}\label{expa}
  \left[{\cal G}^{-1}_{\mu}(\omega)\right]_{\rm s} \simeq  \left[ {\cal G}^{-1}_{\mu}(E_{\mu, \rm s})\right]_{\rm s}  +
  Z_{\mu, \rm s}^{-1} (\omega - E_{\mu, \rm s}),
  \end{equation}
  where we have introduced the notation $\left[{\cal G}^{-1}_{\mu}(\omega)\right]_{\rm s}= \langle {\mu, \rm s} | {\cal G}^{-1}_{\mu}(\omega) | {\mu, \rm s} \rangle$, hence, $\left[{\cal G}^{-1}_{\mu}(E_{\mu, \rm s})\right]_{\rm s}=0$
  and the quasiparticle weight $ Z_{\mu, \rm s}$ such that
  \begin{eqnarray}\label{am}
Z_{\mu, \rm s}^{-1} =  \left[ \frac{\partial {\cal G}_{\mu}^{-1}(\omega)}{\partial \omega} \right]_{\rm s, E_{\mu, \rm s}}.
\end{eqnarray}

  In this way,   the explicit expression for the  subgap Green's function
  in the basis of Shiba states reads,
  \begin{equation}\label{gshi}
{\cal G}_{\mu}(\omega) = \sum_{\rm s}    \; | {\mu, \rm s} \rangle  \frac{Z_{\mu, \rm s}}{\omega - E_{\mu, \rm s} + i \eta } \langle {\mu, \rm s}| ,
\end{equation}
where we have introduced an infinitesimal $\eta >0 $ to regularize the denominator. For a single impurity, we typically have two Shiba states per orbital channel. 

With these definitions, we can write the  spin-resolved spectral density for the particle and hole components of a given Shiba state
 with energy $E_{\mu, \rm s}$ as
\begin{equation}\label{rho}
\rho_{\sigma}^{e,h}({\bf r}, E_{\mu, \rm s})=   \mbox{Tr}\Big[   \Pi^{e,h}_{\sigma} \gamma_{\mu}({\bf r}, E_{\mu, \rm s} )  \;  \rho_{\mu, \rm s}
 \;
\gamma_{\mu}({\bf r}, E_{\mu, \rm s}) \Big], 
\end{equation}
where $\rho_{\mu, \rm s}= \pi Z_{\mu, \rm s} | \mu, \rm s \rangle   \langle \mu, \rm s |$, while $\Pi^{e,h}_{\sigma}$ projects on the subspace of electrons ($e$), holes ($h$) and spin $\sigma$.

In the case of a single impurity embedded in a {\em substrate with a constant density of states} $\nu$ we can get some simple expressions for the energies of the Shiba states. Also, for
vanishing scattering potential, the structure of the Shiba states simplifies and we can also get simple analytical expressions for the quasiparticle weights. 

Using the expression of the matrix $F$, defined in Eq. (\ref{def-f}), and calculated in Appendix \ref{f} for a constant density of states, 
\begin{equation}
F(\omega)= - \frac{\pi \nu }{ \sqrt{\Delta^2-\omega^2}} \sigma_0 \left[ \omega \tau_0 + \Delta \tau_x \right], \;\;\;\;|\omega| < \Delta,
\end{equation}
we can analytically calculate  the energies of the Shiba states from the zeroes of the auxiliary matrix defined in Eq. (\ref{lambda}). This leads to 
\begin{eqnarray}
 & & \mbox{Det}\left[1- F(\omega) V_{\mu}\right]
 = \mbox{Det} \left\{  \tau_0 \sigma_0 + \frac{\pi \nu}{\sqrt{\Delta^2-\omega^2}} \left[ \omega \left( U_{\mu} \tau_z \sigma_0 \right. \right. \right. \nonumber \\ 
& & \left. \left. \left. \;\;\;\;\;\;\;\;\;\;\;\;\; + J_{\mu} S \sigma_z \tau_0 \right)  +
 \Delta \left( i U_{\mu} \tau_y \sigma_0 + J_{\mu} S \tau_x \sigma_z \right) \right] \right\}=0.
 \end{eqnarray}
After simple algebra, we get the energies for the Shiba subgap states,\cite{yang}
\begin{equation}\label{emu}
E_{\mu}^{\pm} = \pm \Delta \frac{ \left(1 + \beta_{\mu}^2-\alpha_{\mu}^2 \right)}{\sqrt{4 \alpha_{\mu}^2 + \left(1- \alpha_{\mu}^2 + \beta_{\mu}^2 \right)^2}},
\end{equation}
being $ |\mu, \pm\rangle$ the corresponding states, with 
\begin{equation}\label{alphabeta}
\alpha_{\mu}= \frac{J_{\mu}}{J_0}, \;\;\;\;\;\;\;J_0=\frac{1}{ \pi \nu S},  \;\;\;\;\;\;\; \beta_{\mu} = U_{\mu} \pi \nu. 
\end{equation}
 If we assume  $U_{\mu}>0$,  the crossing of the two Shiba states, corresponding to $E_{\mu}^{\pm}=0$,
takes place when $J_{\mu}$ achieves the critical value 
\begin{equation}\label{jcmu}
J_{\mu}^{c}= J_0 \sqrt{1+ \beta_{\mu}^2}.
\end{equation}
The corresponding Shiba states  $|\mu, \pm \rangle$  have a very simple form in the case of vanishing potential scattering $U_{\mu}=0$. They read
\begin{eqnarray}\label{shibau0}
|\mu,+\rangle &=& (0,1,0,-1)^T, \;\;\;\;\;\;\;\; |\mu,-\rangle = (1,0,1,0)^T,
\end{eqnarray}
for $J<J_{\mu}^c$, with $\pm$ corresponding to the labeling of Eq. (\ref{emu}). The roles of $\pm$ are interchanged in Eq. (\ref{shibau0}) for $J>J_{\mu}^c$.

The  quasiparticle weights defined in Eq. (\ref{am}) can be calculated from
\begin{eqnarray}
Z_{\mu, {\rm s}}^{-1}
%=\left[ \frac{\partial {\cal G}^{-1}(\omega)}{\partial \omega} \right]_{\omega_{\rm sh}}  
& = &  -  \langle \mu, {\rm s} | V_{\mu} \left[ \frac{\partial F_{\mu}(\omega)}{\partial \omega} \right]_{E_{\mu, {\rm s}} } V_{\mu}  |\mu, {\rm s}\rangle \\
& = &\frac{\nu \pi}{\sqrt{ \Delta^2- E_{\mu, {\rm s}}^2}}
\langle \mu, {\rm s} | V_{\mu} \left[ \tau_0 + \frac{\Delta E_{\mu, {\rm s}}}{\Delta^2 - E_{\mu, {\rm s}}^2} \tau_x \right] V_{\mu}|\mu, {\rm s}\rangle. \nonumber
\end{eqnarray}
This expression significantly simplifies for deep Shiba states which correspond to $|E_{\mu, {\rm s}}| \ll \Delta$,
\begin{equation}\label{zmu}
Z_{\mu, {\rm s}}^{-1}
 \simeq \frac{\nu \pi}{\Delta}
\langle \mu, {\rm s} | V^2_{\mu} |\mu, {\rm s}\rangle. \nonumber
\end{equation}
For vanishing scattering potential, $U_{\mu} \sim 0$, they simplify further and read $Z_{\mu, {\rm s}}^{-1} \simeq \nu \pi (S J_{\mu})^2/\Delta$.

\subsubsection{Several impurities}
We can  easily generalize the  procedure explained before to calculate the energies of the Shiba states in the case of many impurities. As before,  we start by defining the matrix 
\begin{equation}\label{inv}
\hat{\Lambda}(\omega)= \hat{V} - \hat{V} \hat{F}(\omega) \hat{V}
\end{equation}
 and
diagonalize it for every $\omega$. Here, the matrices $\hat{V}$ and $\hat{F}$ enclose all the matrix elements of $V^j_{\mu}$ and $F^{j,j^{\prime}}_{\mu,\mu^{\prime}}(\omega)$.
We denote the corresponding basis of Nambu spinors as $|j,\mu\rangle$.  Then, for each 
$\omega$, we have an expression identical to Eq. (\ref{esp-shiba}), with the Shiba energy defined from the condition that some eigenvalue vanishes. 
There are  $2 N_s$ of such states, being $N_s$ the number of impurities times the number of orbitals.
We name the corresponding Shiba energies $E_{\ell, \rm s}, \; \ell= 1, \ldots, N_{s}$, ${\rm s} =\pm$ and the 
corresponding eigenvectors are $|\ell, \pm \rangle$, with $E_{\ell, +}=-E_{\ell, -}$ and $| \ell, - \rangle ={\rm CT} \; | \ell, +  \rangle$.

The generalization of Eq. (\ref{gshi}) for the case of many impurities reads
\begin{equation}\label{gshin}
\hat{\cal G}_{\ell}(\omega) =   \sum_{\rm s=\pm}   \; | \ell, {\rm s} \rangle  \frac{Z_{\ell,\rm s} }{\omega - E_{\ell, \rm s} + i \eta } \langle \ell, {\rm s} |, 
\end{equation}
%\begin{equation}\label{tman}
%T^{j,j^{\prime}}_{\mu, \mu^{\prime}}(\omega) \simeq  V^{j}_{\mu} \sum_{l}   \frac{ 1 }{A^s_l} \; | \Phi_l \rangle  \frac{1}{\omega - E_l + i \eta } \langle \Phi_l |  
%V^{j^{\prime}}_{\mu^{\prime}}.
%\end{equation}
and the spin-resolved spectral density for the particle and hole components of 
the Shiba states is evaluated in a similar way to the case of a single impurity. For a given state with eigenenergy $E_{\ell, \rm s}$ it reads
\begin{equation}\label{rho2}
\rho_{\sigma}^{e,h}({\bf r}, E_{\ell, \rm s})=  \sum_{\mu, \mu^{\prime} j, j^{\prime}}  \mbox{Tr}\left[   \Pi^{e,h}_{\sigma} \gamma_{\mu}({\bf r}-{\bf r}_j, E_{\ell,\rm s}) \;
\hat{\rho}_{\ell, \rm s} \;
%\rho^{j,j^{\prime}}_{\mu,\mu^{\prime}}(E_l)  
\gamma_{\mu^{\prime}}({\bf r}_{j^{\prime}}- {\bf r}, E_{\ell,\rm s})\right],
%\mbox{Im} \left( T^{j,j^{\prime}}_{\mu,\mu^{\prime}}(E_l) \right)
%\frac{ \left( \sigma_0  \pm \sigma_z \right) }{2} (\tau_0 + \tau_z)  
\end{equation}
with $\hat{\rho}_{\ell, \rm s} = \pi Z_{\ell, \rm s} |\ell, \rm s \rangle \langle \ell, \rm s| $. Here we recall that the Shiba states $| \ell, \rm s \rangle$ are expanded by the Nambu spinors $|j , \mu \rangle$.
%where  we have introduced the following definition
%\begin{eqnarray} 
%\rho^{j,j^{\prime}}_{\mu,\mu^{\prime}}(E_l)  &=& - 2 \mbox{Im} \left( T^{j,j^{\prime}}_{\mu,\mu^{\prime}}(E_l) \right) = -\frac{\pi}{A_l} V_{\mu}^j |\Phi_l \rangle \langle \Phi_l | V_{\mu}^{j^{\prime}}.
%\end{eqnarray}

\subsection{Effective tight-binding - BCS  model for dilute impurities}\label{sec:eff}
The matrix elements $F^{j, j^{\prime}}_{\mu, \mu^{\prime}}(\omega) = F^{j^{\prime},j}_{\mu^{\prime}, \mu}(\omega)$ define an effective interaction between electrons in subgap states  localized at the impurities. 
The strength of this interaction depends on the separation between the impurities $r_{j, j^{\prime}}$ relative to the coherence length of the superconductor $\xi_0$ and to the wavelength $k_F^{-1}$. Typically, 
$r_{j, j^{\prime}} \leq \xi_0$.  
We focus on a situation where the impurities are sufficiently diluted to justify a perturbative treatment of the off-diagonal elements $F^{j, j^{\prime}}_{\mu, \mu^{\prime}}$, with $j \neq j^{\prime}$. This corresponds to distances
satisfying $k_F r_{j, j^{\prime}} \gg 1$. Under these conditions we can solve the equation for the Shiba states starting from the solution of the isolated impurities and
derive an effective Hamiltonian. Our strategy is to get a convenient expression for the inverse of the subgap Green's function defined in Eq. (\ref{inv}) in this limit and then to identify the effective
Hamiltonian $ \hat{H}_{\rm eff}$ from the relation
\begin{equation}\label{gm1}
\hat{\cal G}^{-1}(\omega)= \left(\omega \hat{1} - \hat{H}_{\rm eff}\right).
\end{equation}
where $\hat{1}$ is the identity matrix.
We start by writing $\hat{\cal G}^{-1}(\omega)$ for the $N$-impurity case as follows,
\begin{equation}
\hat{\cal G}^{-1}(\omega) = \sum_{j, \mu}  \left[{\cal G}^{-1}(\omega)\right]_{\mu}^{j} +  \sum_{j,j^{\prime}, \mu, \mu^{\prime}}  \left[{\cal G}^{-1}(\omega)\right]^{j,j^{\prime}}_{\mu, \mu^{\prime}},
\end{equation}
with
\begin{eqnarray}\label{gm1per}
\left[{\cal G}^{-1}(\omega)\right]_{\mu}^{j} &= & V_{\mu}^{j} - V_{\mu}^{j} F_{\mu}^{j,j}(\omega) V_{\mu}^{j}, \nonumber \\
\left[{\cal G}^{-1}(\omega)\right]^{j,j^{\prime}}_{\mu, \mu^{\prime}}&=&V_{\mu}^{j} F_{\mu,\mu^{\prime}}^{j,j^{\prime}}(\omega) V_{\mu}^{j^{\prime}}, \;\;\;\;\;j \neq j^{\prime}.
\end{eqnarray}
For energies close to the Shiba energies for the single impurities,  $E_{j, \mu \rm s}$, we expand $\hat{\cal G}^{-1}(\omega)$ with respect to this value.  Projecting on the states $ | j, \mu,{\rm s} \rangle = 1/\sqrt{Z_{j, \mu, \rm s}} |\mu, {\rm s} \rangle^{j}$, where $|\mu, {\rm s} \rangle^{j}$ are the Shiba states of the isolated impurity $j$,
 and considering the inter-impurity terms as perturbations, we have
\begin{eqnarray}
\left[{\cal G}^{-1}(\omega)\right]_{\mu}^{j} &\simeq  & \sum_{\rm s} |  j, {\mu, \rm s} \rangle \left( \omega- E_{j, \mu,\rm s} \right) \langle j, \mu, {\rm s} |, \\
\left[{\cal G}^{-1}(\omega)\right]^{j,j^{\prime}}_{\mu, \mu^{\prime}}& \simeq & \sum_{{\rm s}, {\rm s}^{\prime}} |  j, {\mu, \rm s} \rangle \;
h_{\mu s, \mu^{\prime} s^{\prime} }^{j, j^{\prime}}(\omega)\;
% \sqrt{Z_{j, \mu, \rm s}  } V_{\mu}^{j} F_{\mu,\mu^{\prime}}^{j,j^{\prime}}(\omega) V_{\mu^{\prime}}^{j^{\prime}}   \sqrt{ Z_{ j^{\prime}, \mu^{\prime},{\rm s}^{\prime} } } 
 \langle j^{\prime}, \mu^{\prime}, {\rm s}^{\prime} |,\nonumber 
%\hat{\cal G}^{-1}(\omega)  & \simeq  &  \sum_{j,j^{\prime}, \mu, \mu^{\prime}, {\rm s}, {\rm s}^{\prime}} |  j, {\mu, \rm s} \rangle \left[ \frac{\left( \omega- E_{\mu,\rm s} \right)}{Z_{\mu, \rm s}} \delta_{j,j^{\prime}} 
%\delta_{\mu,\mu^{\prime}} 
%\delta_{{\rm s},{\rm  s}^{\prime}} \right. \nonumber \\
%& & \left.  +
%V_{\mu}^{j} F_{\mu,\mu^{\prime}}^{j,j^{\prime}}(\omega) V_{\mu^{\prime}}^{j^{\prime}} \right] \langle j^{\prime}, \mu^{\prime}, {\rm s}^{\prime} |.
\end{eqnarray}
being
\begin{equation} \label{fmumu}
h_{\mu s, \mu^{\prime} s^{\prime} }^{j, j^{\prime}}(\omega) =\sqrt{Z_{j, \mu, \rm s} Z_{ j^{\prime}, \mu^{\prime},{\rm s}^{\prime} } }  \langle  j, \mu,{\rm s} |V_{\mu}^j F_{\mu, \mu^{\prime}}^{j, j^{\prime}}(\omega) V_{\mu^{\prime}}^{j^{\prime}}   | j^{\prime}, \mu^{\prime}, {\rm s}^{\prime} \rangle.
\end{equation}
If we assume that the Shiba states are deep in energy, $E_{\mu, \rm s} \sim0$, we can approximate $h_{\mu s, \mu^{\prime} s^{\prime} }^{j, j^{\prime}}(\omega) \sim h_{\mu s, \mu^{\prime} s^{\prime} }^{j, j^{\prime}}(0)$. We have verified in the explicit calculations that this approximation also works for high-energy Shiba states with $E_{\mu, \rm s} \sim \Delta$, since this function depends mildly on $\omega$ for sufficiently distant impurities.

Thus, the effective Hamiltonian  in the basis of Shiba states of the isolated impurities reads
\begin{eqnarray} \label{hefjjp}
\left[ \hat{H}_{\rm eff} \right]_{j,j}  & = & \sum_{\mu,s} | j, {\mu, s} \rangle  \; E_{j, \mu, s} \; \langle  j, \mu,s | ,  \\
 \left[  \hat{H}_{\rm eff} \right]_{j,j^{\prime}}  & = &  
\sum_{s,s^{\prime},\mu,\mu^{\prime}} | j, \mu, {\rm s} \rangle \;  h_{\mu s, \mu^{\prime} s^{\prime} }^{j, j^{\prime}}(0)\;
 \langle  j^{\prime}, \mu^{\prime}, {\rm s}^{\prime} |,\;\;  j \neq j^{\prime},\nonumber
\end{eqnarray}
where the effective inter-impurity interaction is given by the second line of Eq. (\ref{hefjjp}).  

In order to get analytical expressions for the matrix elements, we introduce some simplifying assumptions. In particular, we focus on the case where the scattering potentials $U_{\mu}$ can be neglected, in 
which case, the Shiba states can be simply expressed as in Eq. (\ref{shibau0}). Notice, however, that the latter are expressed in the quantization axis for the spin oriented along the magnetic moment of the impurity.
In general, the impurities have different orientations  $\hat{\bf S}_j = {\bf S}_j/S=\left( \sin \theta_j \cos \phi_j,\sin \theta_j \sin \phi_j , \cos \theta_j \right)$. Hence, the Shiba states localized at the impurity $j$
read
 \begin{eqnarray}\label{states}
|j,\mu,+\rangle &=& \left( |j + \rangle,  |\overline{j + }\rangle \right)^T,\nonumber \\
 |j,\mu,-\rangle & = & \left( |j - \rangle,  |\overline{j -} \rangle \right)^T,
 % | j, \mu, +  \rangle = |+ \tau_x \rangle |j \uparrow \rangle, \;\;\;\;\;\;\;\;\;\;  | j, \mu, -  \rangle = |- \tau_x \rangle |j \downarrow \rangle.
 \end{eqnarray}
with $|\overline{j \pm }\rangle ={\rm T} |j \pm \rangle$ and
 \begin{eqnarray} \label{trans}
 |j - \rangle & = & \left( \cos(\theta_j/2), \; \sin(\theta_j /2) e^{i \phi_j} \right)^T,  \nonumber \\
 |j + \rangle & = & \left( -\sin(\theta_j/2) e^{-i \phi_j}, \; \cos(\theta_j /2)  \right)^T.
 \end{eqnarray}

 This leads to the following structure for the  effective Hamiltonian
 \begin{equation}\label{heff}
 {\cal H}_{\rm eff} = \left(\begin{array}{cc} {\cal H}_{\mu, \mu^{\prime} }^{j j^{\prime}} & \Delta_{\mu, \mu^{\prime} }^{j j^{\prime}} \\
( \Delta_{\mu, \mu^{\prime} }^{j j^{\prime}})^{\dagger} & - ({\cal H}_{\mu, \mu^{\prime} }^{j j^{\prime}})^* \end{array} \right).
 \end{equation}
 The diagonal matrix elements are $h^{j j}_{\mu, \mu } =E^j_{\mu, +} $, while the other matrix elements have the form
 \begin{equation}\label{heod}
 {\cal  H}_{\mu, \mu^{\prime} }^{j j^{\prime}} = \eta_{\mu, \mu^{\prime} }^{j j^{\prime}}  \langle j + |  j^{\prime} + \rangle, \;\;\;\;\;\;\;\;\;\;  \Delta_{\mu, \mu^{\prime} }^{j j^{\prime}} = 
  \delta_{\mu, \mu^{\prime}}^{j j^{\prime}}  \langle j +  |  j^{\prime} - \rangle, \;\;\; j\neq j^{\prime}.
 \end{equation}
 Explicit expressions for the $\eta_{\mu, \mu^{\prime} }^{j j^{\prime}}$ and $ \delta_{\mu, \mu^{\prime}}^{j j^{\prime}} $ for the  case of a substrate modeled by plane waves will be presented in the next Section.
 
 The spectrum of Shiba states in this regime is given by the eigenenergies of ${\hat H}_{\rm eff}$, which we name them  $\tilde{E}_{\ell, \rm s}$ and $| \ell, \rm s \rangle_{\rm eff}$ are the corresponding eigenstates. 
 These states can  be used to calculate the spectral density in the dilute regime. The corresponding expression reads
\begin{equation}\label{rhoeff}
\rho_{\rm eff, \sigma}^{e,h}({\bf r}, \tilde{E}_{\ell, \rm s})=  \sum_{\mu, \mu^{\prime} j, j^{\prime}}  \mbox{Tr}\left[   \Pi^{e,h}_{\sigma} \gamma_{\mu}({\bf r}-{\bf r}_j, \tilde{E}_{\ell,\rm s}) \;
 %|\Phi_l \rangle \langle \Phi_l | \; 
 %\rho^{j,j^{\prime}}_{\mu,\mu^{\prime}}(E_\ell)  
 \hat{\rho}^{\rm eff}_{\ell, \rm s}
\gamma_{\mu^{\prime}}({\bf r}_{j^{\prime}}- {\bf r}, \tilde{E}_{\ell,\rm s})\right], 
%\mbox{Im} \left( T^{j,j^{\prime}}_{\mu,\mu^{\prime}}(E_l) \right)
%\frac{ \left( \sigma_0  \pm \sigma_z \right) }{2} (\tau_0 + \tau_z)  
\end{equation}
where $\hat{\rho}^{\rm eff}_{\ell, \rm s}$ is the density operator associated to the eigenstates of the effective Hamiltonian expanded in the basis of the Nambu spinors
$|j, \mu \rangle$.

 \section{Results}\label{sec:res}
 
 \subsection{Heuristic description of the YSR states in single impurities and  dimers}\label{sec:heur}
As an introduction to the formal analysis based on the numerical solution of the subgap Green's function and the effective Hamiltonian presented before, we find it convenient to start by providing some qualitative arguments. We
review the quantum phase transition  in a single-orbital  classical impurity and infer on the basis of simple arguments, how this picture  generalizes in the mutiorbital case.
We also extend our analysis to the case of a dimer. 

\begin{figure}[t]\begin{center}
  \includegraphics[width=\columnwidth,height=4.2cm]{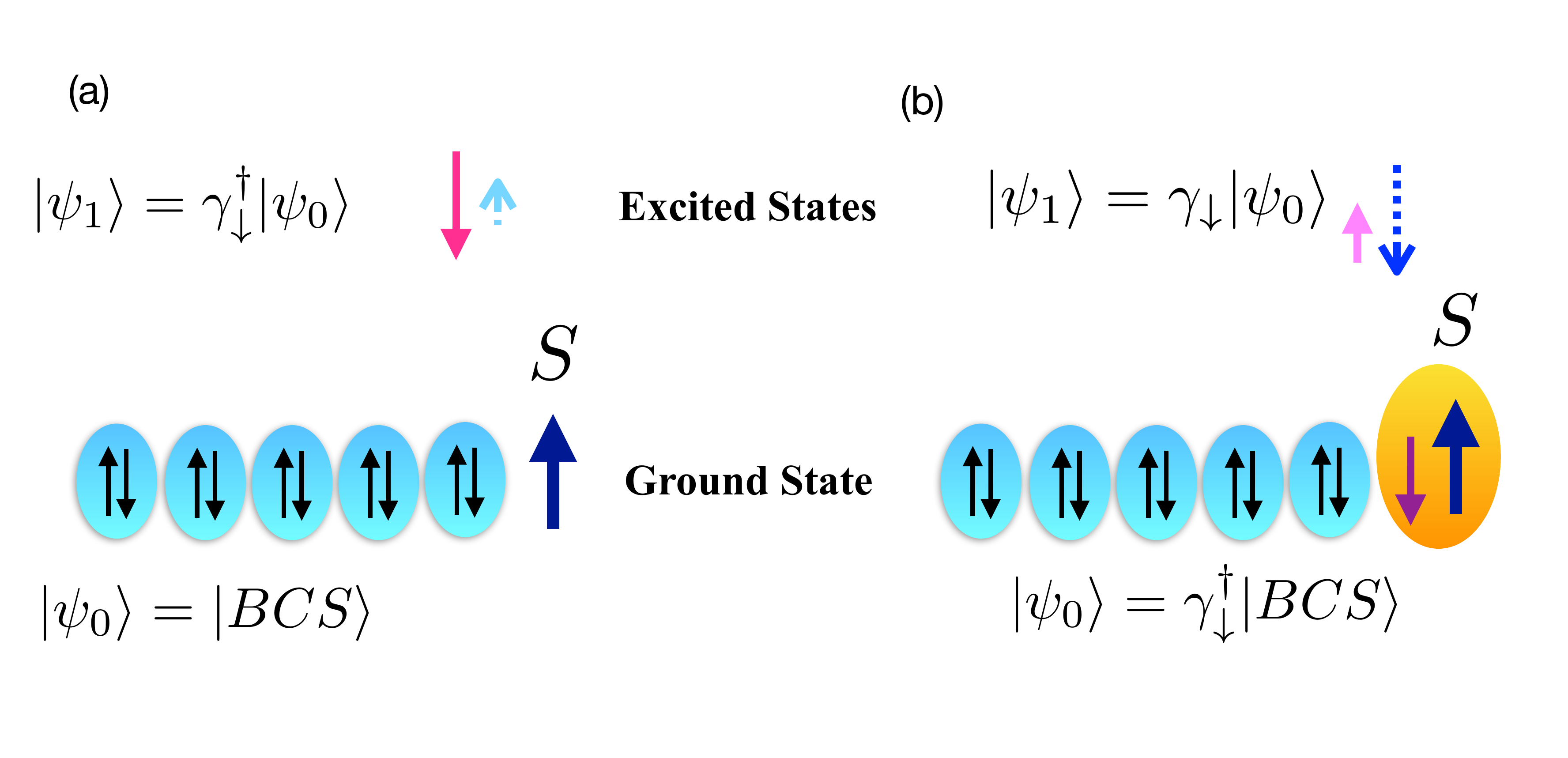}
    \caption{Sketch of the expected scenarios for the ground state and excited states with positive energy of a single-orbital impurity with spin $S$  antiferromagnetically coupled to a BCS substrate.
        (a) $J < J^c$. The ground state is a free impurity in the BCS condensate of the substrate. Subgap excited states correspond to quasiparticles with electron  and hole components with the electron component oriented 
    antiparallel to the impurity.  (b) $J > J^c$. The ground state changes parity and consists of a BCS condensate plus a quasiparticle bounded to the impurity. The excited state corresponds to annihilating  the bounded quasiparticle  of the ground state.
     Particle and hole components of the excitations are, respectively, represented by solid and dashed arrows. The dominant component for positive scattering potential  is represented in dark color and larger size of the arrow.}
 \label{fig0}
\end{center}\end{figure}

 Fig. \ref{fig0} shows sketches of the expected scenarios for a single-orbital impurity.
 Assuming antiferromagnetic coupling between the impurity and the superconductor, for $J<J^c$ the electronic ground state is  formed of Cooper pairs, $|\psi_0\rangle =|BCS\rangle$ and the subgap 
 excitation with positive energy corresponds to creating a quasiparticle with spin antiparallel to the impurity 
 on the ground state $|\psi_1 \rangle= \gamma^{\dagger}_{\downarrow} |\psi_0\rangle$. Instead,  for $J>J^c$,  the quasiparticle gets bound to the impurity and the 
 roles of the states $|\psi_0\rangle$ and $|\psi_1\rangle$ are interchanged,  $|\psi_1\rangle$ becoming the ground state. The excited state
 corresponds to annihilating the bound quasiparticle. Each of these states have weights on particle and hole states of the free-electron basis. 
 For a finite scattering potential  interaction $U>0$, these weights are different. 
  
  \begin{figure}[t]\begin{center}
  \includegraphics[width=\columnwidth,height=7cm]{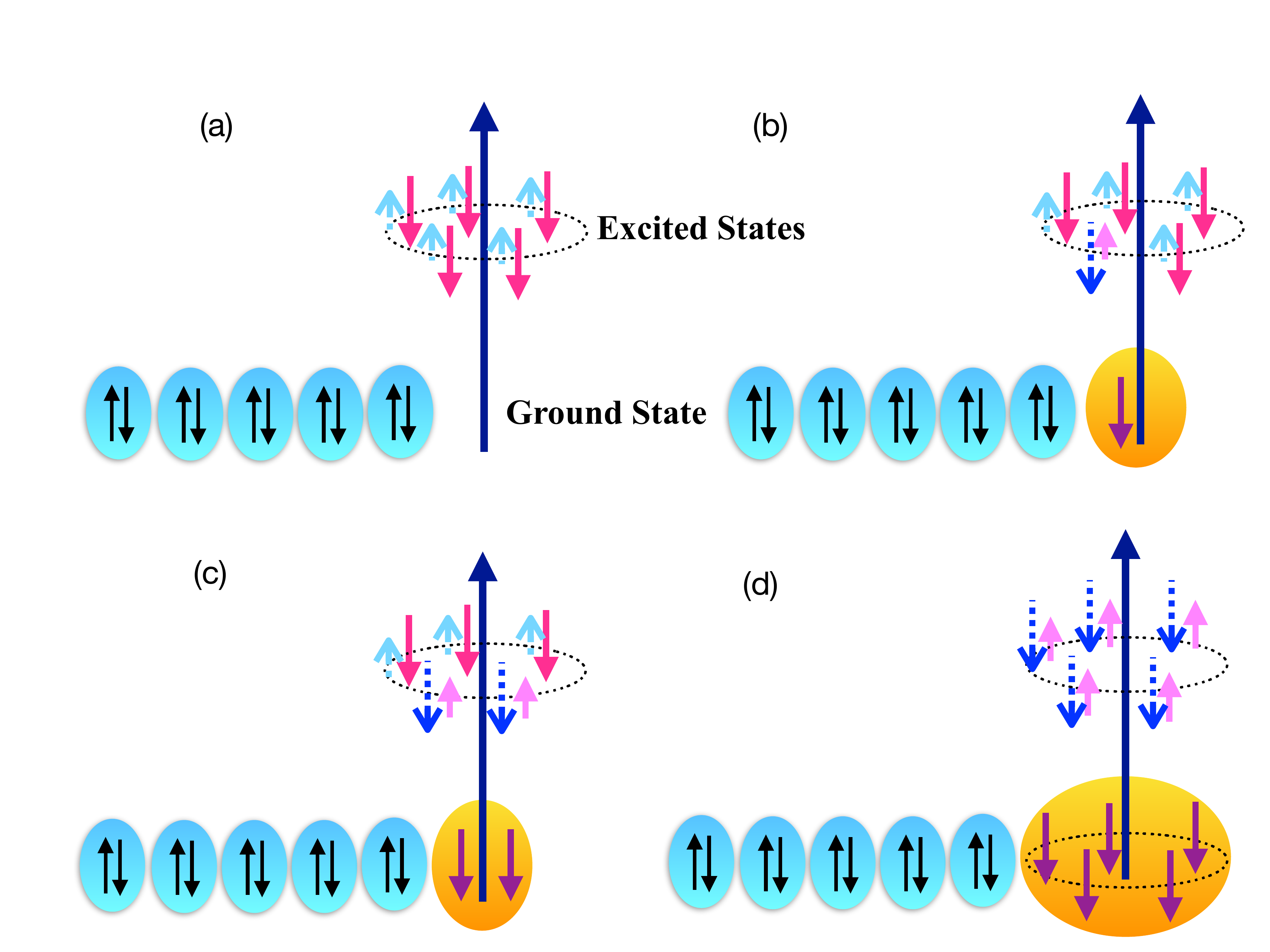}
    \caption{Sketch of the expected scenarios for the ground state and excited states with positive energy of an impurity with several orbitals placed in a superconducting substrate with crystal-field splitting.
    Particle and hole components of the excitations are, respectively, represented by solid and dashed arrows. The dominant component is represented in dark color and larger size of the arrow.
    (a) $J_{\mu} < J_{\mu}^c, \; \forall \mu$. The impurity is completely unscreened. The ground state consists of a BCS condensate and a free magnetic moment. (b) One of the channels is strongly coupled with 
    $J_{\nu} > J_{\mu}^c$  while the others have $J_{\mu} < J_{\mu}^c,\; \mu\neq \nu$.  The ground state is  a BCS condensate plus a quasiparticle bounded to the impurity.  
    (c) Two channels are strongly coupled and the others are weakly coupled. 
    (d) Quasiparticles in all the available channels are bounded to the impurity. The ground state has total spin zero and the impurity is completely screened.}
 \label{fig01}
\end{center}\end{figure}

  For a multiorbital impurity with  weak $J_{\mu}$ compared to $\Delta$, the ground state corresponds to the free impurity in the BCS condensate of the substrate, while the YSR states correspond to quasiparticle  subgap excitations with electron component antiparallel to the impurity in all the orbital channels. This situation is illustrated in Fig. \ref{fig01} (a). In the opposite limit, when the exchange couplings overcome a critical value $J_{\mu}^c$ given by Eq. (\ref{jcmu}), quasiparticles in
  all the channels get bound to the impurity and the ground state corresponds to the BCS condensate of the substrate plus the $2l+1$ bound quasiparticles antiparallel to the impurity. This is  illustrated in
  Fig. \ref{fig01} (d) and it is akin to the quantum phase transition of the single orbital case. The difference consists in the fact that the change of parity in the ground states takes place in the different orbital channels. 
  Making an analogy to the Kondo problem for quantum impurities, the picture of Fig.  \ref{fig01} (a) corresponds to the unscreened impurity, while that of Fig.  \ref{fig01} (d) corresponds to the fully screened impurity.
  The regime analogous to the underscreened Kondo effect of the quantum impurity corresponds to a ground state having bound quasiparticles only in
  some of the orbital channels, as illustrated in Figs.  \ref{fig01} (b) and (c).

 \begin{figure}[t]\begin{center}
  \includegraphics[width=\columnwidth,height=5cm]{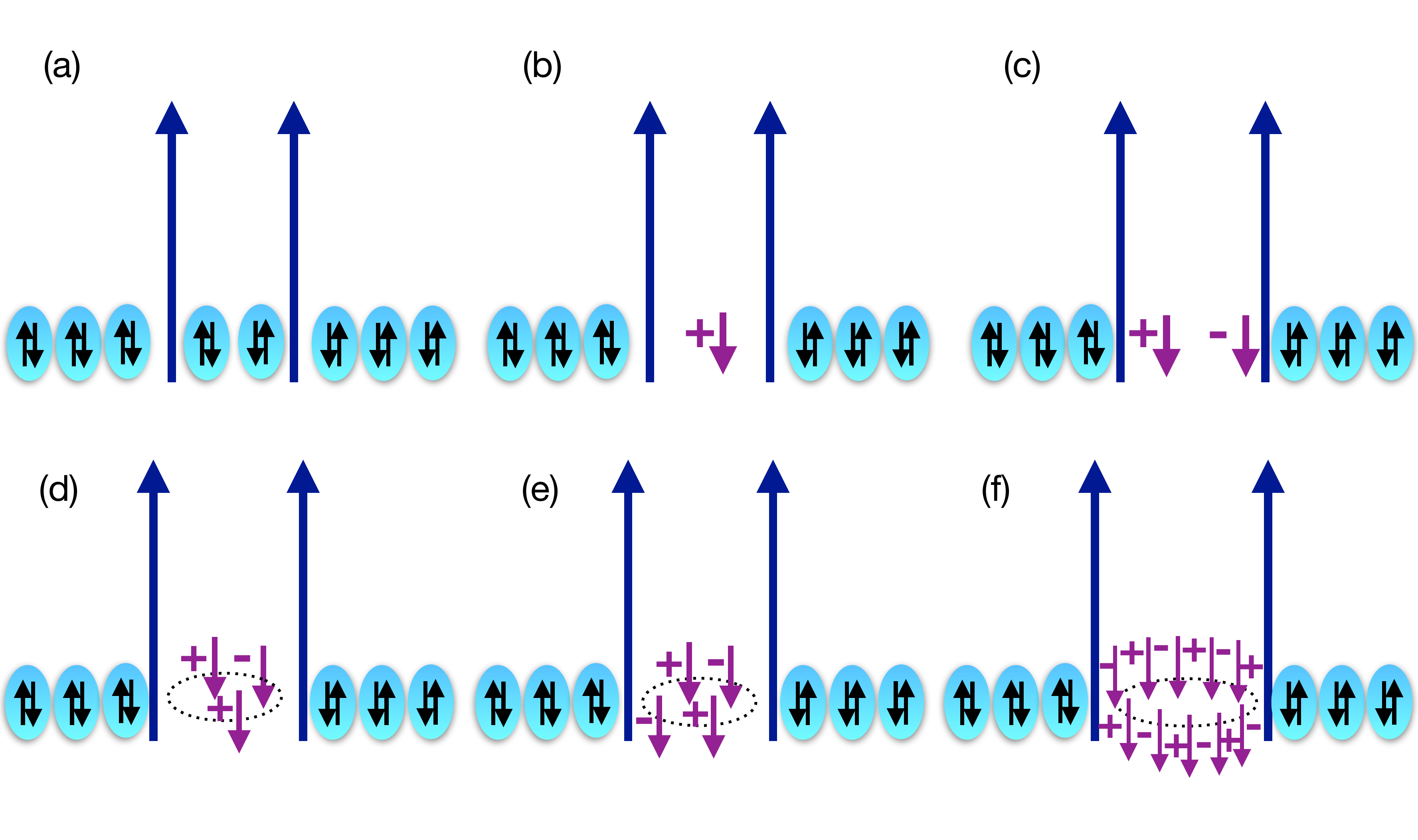}
    \caption{Sketch of the expected scenarios for the ground state of two impurities  with several orbitals placed in a superconducting substrate with crystal-field splitting with the net spin aligned
    in the same direction and separated by a distance  $r> k_F^{-1}$. The YSR states form in bonding $(+)$ and antibonding $(-)$ combinations of states localized at the impurities.
    (a) The impurities are completely unscreened. The ground state has total spin $2S$. (b) The bonding combination of one of the angular-momentum channels is bounded, while the others become 
    unbounded.  (c) (d) (e) The impurities are partially screened by bonding and/or antibonding states in different channels.
    (f) Bonding and antibonding states in all the channels are bounded and the impurity is fully screened.}
     \label{fig02p}
\end{center}\end{figure}

 For dimers, the scenario depends crucially on the relative alignment of the magnetic moments. Intuitively, we can expect to easily generalize the scenario of a single impurity in the case of two with parallel magnetic moments.
 Here it is also important to take into account the  inter-impurity distance $r$, relative to the localization length of the YSR of the single impurity. As mentioned before, there are two relevant length scales in the problem, which set the localization length of the YSR states of the 
 single impurity. One is the Fermi wave length $\lambda_F= 2 \pi/k_F$, 
 and the other is the superconducting coherence length $\xi_0$  for the electrons in the substrate. The usual case in experiments is  $\xi_0 \gg \lambda_F$. \cite{PGvO} For $r \ll \lambda_F$, the dimer behaves as
 an effective impurity with magnetic moment $2S$, while in the limit $r \gg \lambda_F$ its behaves as two independent single impurities. For intermediate distances, the YSR states associated to the single impurity
 hybridize forming bonding $(+)$ and antibonding $(-)$ combinations  in every
 orbital channel.  Furthermore, depending on the symmetry of the substrate and of the orbitals, it is also posible that two or more orbital channels hybridize. 
 Depending on the strength of the exchange interaction, none, some, or all of these combined states may cross zero energy, leading to several  quantum phase transitions, in which the impurities can be unscreened, underscreened or fully screened. This  bears resemblance to the scenario of the single impurity as  illustrated in Fig. \ref{fig02p}.

  \begin{figure}[t]\begin{center}
  \includegraphics[width=\columnwidth,height=5cm]{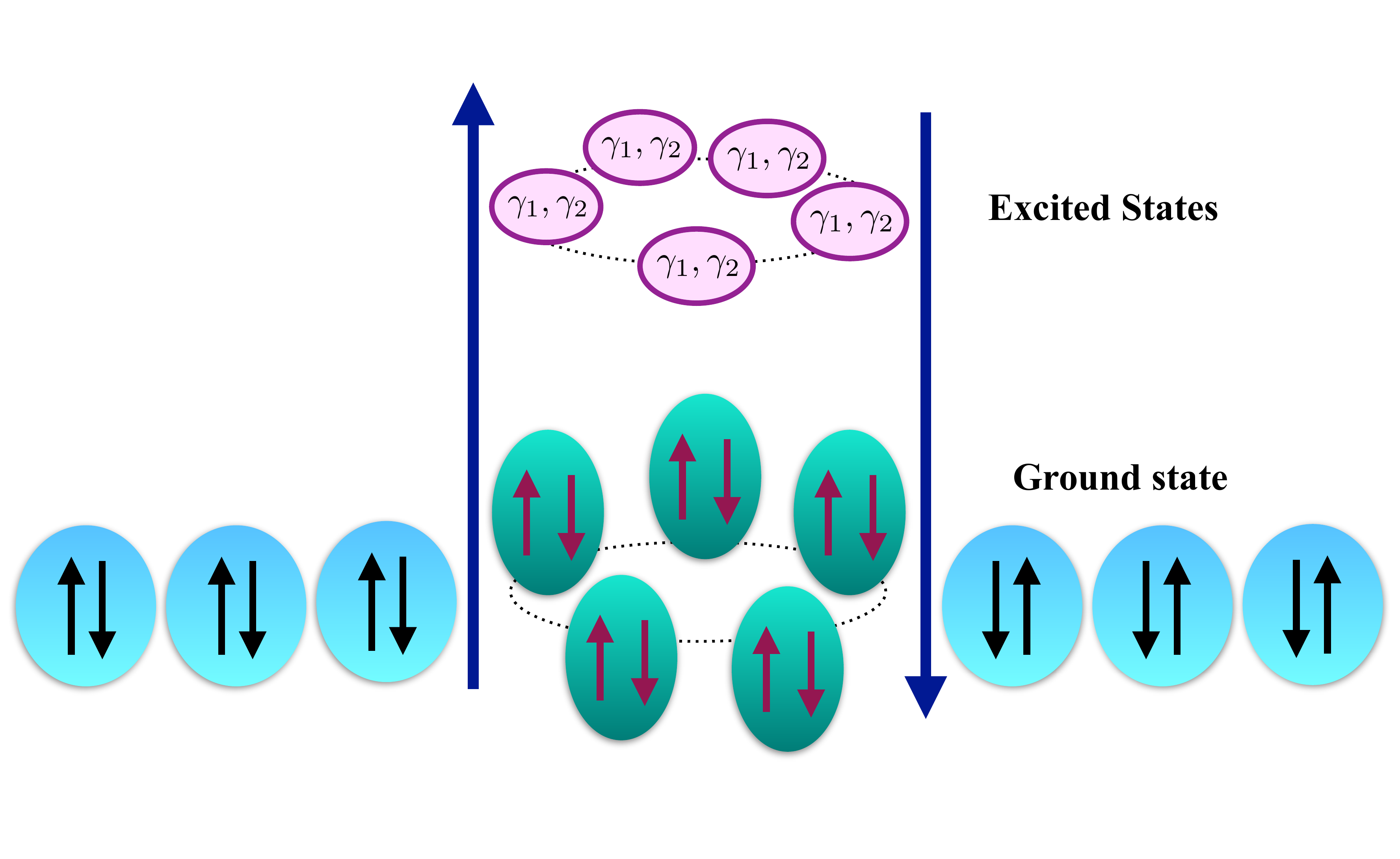}
    \caption{Sketch of the ground state  and excited states of two impurities  with several orbitals placed in a superconducting substrate with crystal-field splitting with the magnetic moments aligned
    in opposite directions and separated by a distance  $r> \lambda_F$. The YSR states  are paired and the excitations, represented with $\gamma_1, \gamma_2$ in every channel,  are degenerate.}
     \label{fig02a}
\end{center}\end{figure}

 For dimers of identical impurities with  magnetic moments aligned in opposite directions the scenario is very different. For $r>\lambda_F$, the 
  YSR states of the individual impurities are coupled only through a pairing interaction. Depending on the symmetry this can be intra-channel or inter-cannel.
  The ground state is the BCS condensate of the substrate plus the BCS ground state of the paired YSR states. The subgap excitations 
  are the Bogoliubov  excitations  of the latter BCS state.   This is illustrated in Fig. \ref{fig02a}.

 \begin{figure}[t]\begin{center}
  \includegraphics[width=\columnwidth,height=5cm]{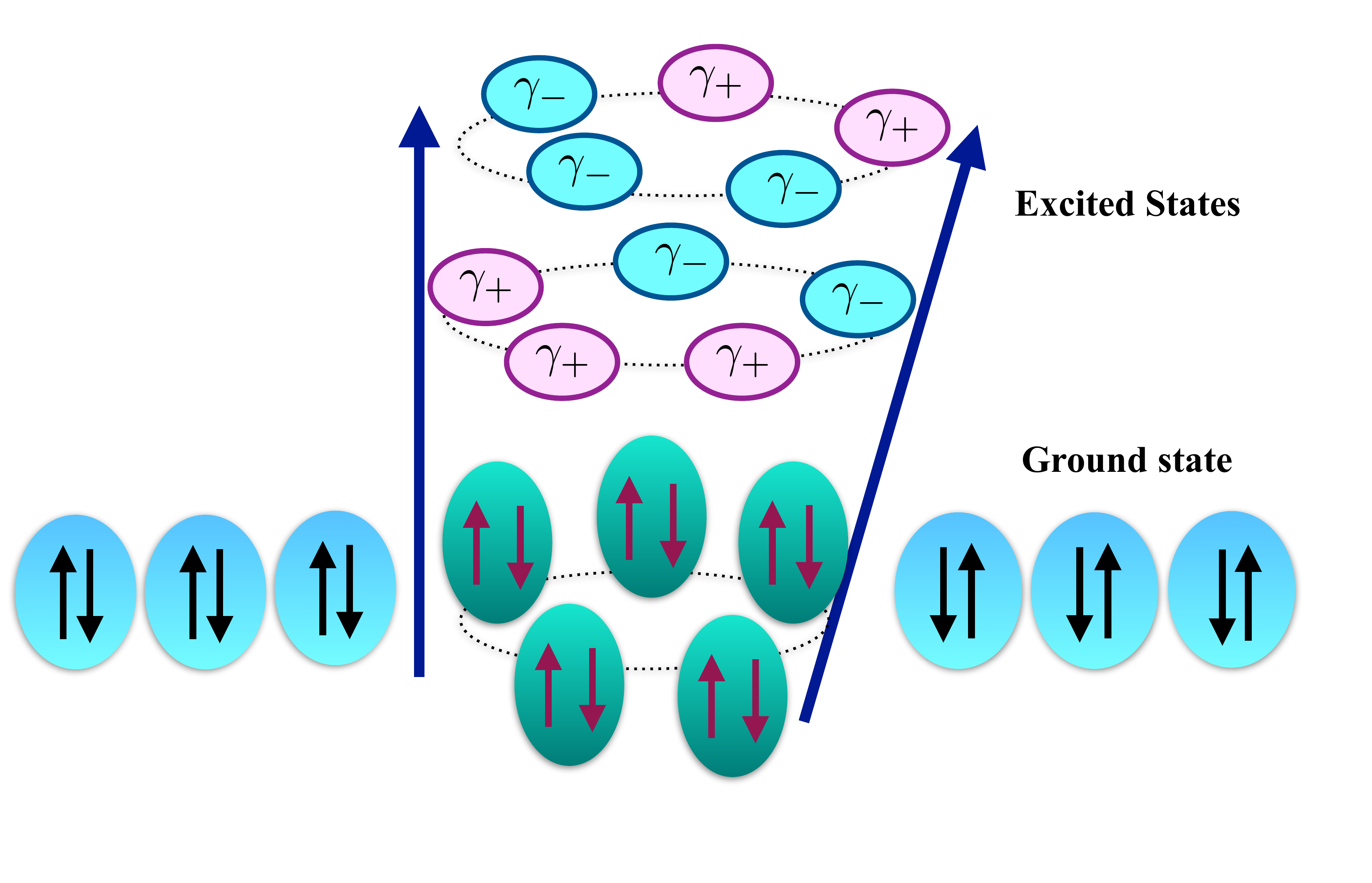}
    \caption{Sketch of the ground state and excited states  of two impurities  with several orbitals placed in a superconducting substrate with crystal-field splitting with the magnetic moments aligned forming an angle $\theta \neq 0, \pi$
    in opposite directions and separated by a distance  $r> \lambda_F$. The excitations disperse in symmetric and antisymmetric combinations represented by $\gamma_+,\gamma_-$.}
     \label{fig03}
\end{center}\end{figure}

The intermediate situation, where the magnetic moments of the impurities are oriented forming a relative angle $\theta \neq 0, \pi$, is illustrated in Fig. \ref{fig03}. In this case there is a combination of a normal hybridization and pairing of the
YSR states of the individual impurities. The ground state is similar to that of the antiparallel states and consists of a BCS condensate of the substrate plus the BCS state of the pairs of YSR of the individual impurities. The excited states are
Bogoliubov excitations of the latter BCS state but in the present case, they disperse as a consequence of the normal hybridization, forming symmetric  and antisymmetric  combinations. In the present configuration of the magnetic moments, time-reversal symmetry is broken. Eventually, some of these quasiparticle excitations may cross zero energy, changing the parity of the ground state. This feature is similar to the case of parallel impurities, although the nature 
of the quasiparticles is different in the two cases. 
This is the scenario favoring the  topological phase in long chains of adatoms. \cite{nadj1,nadj2,PGvO}

\subsection{Numerical results}
 We present results for the case  a  substrate modeled by plane waves with crystal field splitting, focusing on impurities with atomic orbitals $d_{z^2}, d_{x^2-y^2}, d_{xy}, d_{xz}, d_{yz}$ and spin $S=5/2$.
 We will focus on the case of vanishing potential scattering and assuming the same strength of the exchange interaction for all the impurities,
 $V_{\mu}^j = J_{\mu} \; \boldsymbol{\sigma}  \cdot {\bf S}_j$, with ${\bf S}_j = S ( \sin \theta_j \cos \phi_j, \; \sin \theta_j \sin \phi_j, \; \cos \theta_j )$. We also consider that impurities separated by a distance $r$.

 The evaluation of the matrix elements of $F$ defined in Eq. (\ref{fjjp}) in a plane-wave substrate with a constant density of states is presented in Appendix \ref{f} for energies within the gap, $|\omega| \leq \Delta$. 
 The results depend strongly on the nature of the substrate and the different cases can be 
  summarized as follows
  \begin{eqnarray} \label{fjj}
  F^{j,j}(\omega) & = &  - \frac{\nu \pi}{\sqrt{\Delta^2 -\omega^2}} \sigma_0 \left[\omega \tau_0 + \Delta \tau_x \right], \nonumber \\
   F^{j\neq j^{\prime}}_{\mu, \mu^{\prime}}(\omega) & = & - 4 \pi \nu \sum_{l} (i)^{l}  \; {\cal C}_{\mu, \mu^{\prime}} (l)   \left[ \left( \omega \tau_0 + \Delta \tau_x \right) I^{(l)}_1(r_{j, j^{\prime}},\omega) \right. \nonumber \\
   & & \left.
  + \tau_z I^{(l)}_2(r_{j, j^{\prime}},\omega) \right].
  \nonumber\\
  \end{eqnarray}
  The coefficients  ${\cal C}_{\mu,\mu^{\prime}} (l)$ are combinations
 of spherical harmonics and are specified in Eq. (\ref{coef}).
  The functions $I^{(l)}_{1,2}(r)$, defined in Eq. (\ref{ilj})  involve integrals of spherical Bessel functions and integer Bessel functions, which are evaluated in Appendix \ref{integrals}.
  Besides exponentially decaying factors $\propto \exp \left( - \sqrt{ 1- (\omega/\Delta)^2}\; r/\xi_0 \right)$, being $\xi_0=v_F/\Delta$, these functions also decay with the distance as powers of $1/(k_Fr)$.

 The scattering potential does not play a crucial role in introducing different energy scales for the different channels in the case of clusters. This is because the interesting
 physics of clusters is dominated by the 
effective interaction between impurities, mediated by the substrate. The latter behaves differently in the different channels $\mu$. This is determined by the inter-impurity matrix elements 
given in Eq. (\ref{fjj}). In general, they have inter-channel components, in addition to the intra-channel ones. In practice, several of these inter-channel components vanish because of symmetry reasons.
With Eqs. (\ref{fjj}) we can calculate the exact spectrum of YSR states following Section \ref{sec:formal}.

In order to calculate the effective Hamiltonian in the dilute limit,  where the distance $r \gg \lambda_F$, we need the function $h_{\mu,\mu^{\prime}}$ defined in Eq. (\ref{fmumu}).
 From the calculations  presented in Appendix \ref{f}  we get
\begin{eqnarray} \label{hef}
& & h_{\mu s, \mu^{\prime} s^{\prime} }^{j, j^{\prime}}(0)  =   - \frac{4 \pi \Delta}{(JS)^2 } e^{-r/\xi_0} \sum_{l} (i)^{l}  \; 
 {\cal C}_{\mu,\mu^{\prime}} (l)   \\
%\right. & & 
 %\left. 
& & \;\;\;\;\;\; \times  \langle  j, \mu^{\prime},{\rm s}  | V_{\mu^{\prime}}^j    
\left[  \tau_x   \kappa^{(l)}_{1}(k_F r)  + \tau_z 
\kappa^{(l)}_{2}(k_F r) \right] V_{\mu^{\prime}}^{j^{\prime}} | j^{\prime}, \mu^{\prime}, {\rm s}^{\prime} \rangle.\nonumber 
\end{eqnarray}
 The
 functions  $\kappa^{(l)}_{1,2}(k_F r)$ are defined in Eqs. (\ref{kappa0}), (\ref{kappa2}) and (\ref{kappa4}).  

The coefficients $ \eta_{\mu, \mu^{\prime} }^{j j^{\prime}}$ and $\delta_{\mu}^{j j^{\prime}}$  introduced in Eqs. (\ref{heod}), can be calculated 
  Eqs. (\ref{fmumu}) and (\ref{hef}), using Eq. (\ref{states}).
  The result is
 \begin{eqnarray}\label{heodpw}
\eta_{\mu, \mu^{\prime} }^{j j^{\prime}} &=& - 4 \pi \Delta  e^{-r/\xi_0}  \sum_{l} (i)^{l}  \; {\cal C}_{\mu, \mu^{\prime} } (l)
  \kappa^{(l)}_{1}(k_F r),   \nonumber \\
 \delta_{\mu, \mu^{\prime} }^{j j^{\prime}} & = &   4 \pi \Delta e^{-r/\xi_0}  \sum_{l} (i)^{l}  \; {\cal C}_{\mu, \mu^{\prime} } (l) 
\kappa^{(l)}_{2}(k_F r).
 \end{eqnarray}
 Following Ref. \onlinecite{PGvO} we can 
 explicitly calculate the scalar products and perform a gauge transformation to eliminate irrelevant phases. The result is 
  \begin{eqnarray} \label{ef}
  \langle j + |  j^{\prime} + \rangle &= & \cos \frac{\theta_j}{2}  \cos \frac{\theta_{j^{\prime}}}{2}
   e^{ i( \frac{\phi_j-\phi_{j^{\prime}}}{2} )} + \sin \frac{\theta_j}{2} 
    \sin \frac{\theta_{j^{\prime}}}{2} e^{-i( \frac{\phi_j-\phi_{j^{\prime}}}{2})} \nonumber \\
 \langle j -  |  j^{\prime} -  \rangle&=& \cos \frac{\theta_j}{2}  
 \sin \frac{\theta_{j^{\prime}}}{2} e^{ i(\frac{\phi_j-\phi_{j^{\prime}}}{2})} - \sin \frac{\theta_j}{2} 
  \cos \frac{\theta_{j^{\prime}}}{2} e^{- i(\frac{\phi_j-\phi_{j^{\prime}}}{2})}.  \nonumber
 \end{eqnarray}
For the case of a single orbital, this Hamiltonian is  the same as that derived in Ref. \onlinecite{PGvO}.  In our case  we have several orbital components $\mu$.  
The dependence with the distance between the impurities enters through the functions
$\propto \exp\{- r/\xi_0 \}$ as well as the functions $\kappa_{1,2}^{(l)}(r)$, which are products of an oscillating function and a modulating function of powers of $1/(k_F r)$.
 We see that for impurities closer than a Fermi wave length ($r < k_F^{-1}$) the effective interactions introduced by the exact off-diagonal elements of $F$ are large, which implies that the description provided by the effective Hamiltonian is no longer valid.

The explicit evaluation of the spectral density $\rho^{e,h}(x,y,z)$ from Eqs. (\ref{rho}) and (\ref{rho2})  implies the evaluation of the matrices $\gamma_{\mu}$.  The latter is explained in Appendix \ref{gamma}.

 \subsubsection{Dimer with parallel spins}
 We consider the magnetic moments of the two impurities aligned along the $z$ axis.
 As mentioned before, the inter-impurity coupling mediated by the substrate has different components in the different orbital channels and even inter-channel components. This ingredient is enough to lead to non-trivial effects, like the
splitting of the different channels to screen the impurity. The hybridization takes place intra and inter-channel. The latter depends on the symmetry of the orbitals  and the nature of the substrate. For the substrate 
 we are considering here, only the channels $(xz,\;yz)$ hybridize, while the channels  $x^2-y^2$, $z^2$,  $xy$ decouple one-another. This can be explicitly seen by examining the coefficients 
 ${\cal C}_{\mu,\mu^{\prime}}(l)$ given in 
 Eq. (\ref{coef}). 
   
 It is useful to have in mind the physics of a single impurity discussed  previously as a reference of our analysis. As in the case of a single impurity, for a cluster of impurities with the spins aligned in the same direction,
 we can characterize the induced YSR states by the component $S_z$  of the spin. When the two impurities are very far apart, the spectrum of subgap states coincide with that of the two isolated impurities. This is represented by the sketch of Fig. \ref{fig02p} (a). For the assumed parameters, the
 critical coupling is the same in all the channels and equal to $J_c=J_0$ defined in Eq. (\ref{alphabeta}). 
 If the 
 exchange coupling is $J< J_c$, the impurities are  completely unscreened and for $J \geq J_c$ they are screened.
 As the impurities 
 become closer, the YSR states hybridize forming bonding ($+$) and antibonding ($-$) combinations of polarized states antiferromagnetically aligned with the impurities (for  $J< J_c$) and
 in the direction of the polarization of the impurities (for  $J \geq J_c$). The degree of hybridization is different for the different channels, as explained before. For $J>J_c$, it may occur within 
 a given channel,  that the state corresponding to one of the combinations 
 (bonding or antibonding) has a low enough energy to cross zero and get bounded to the impurities. This situation is represented in Fig. \ref{fig02p} (b). For single-orbital impurities, this corresponds to a phase transition to a molecular doublet 
 state, as discussed by Moca et al. \cite{moca} Depending on the strength of the exchange coupling and the inter-impurity distance, several intermediate situations, corresponding to partial screening may take place. Some
 of the possibilities are sketched in Fig. \ref{fig02p} (c) (d) (e). The extreme phase, where all the states are bounded in all the channels corresponds to full screening of the two impurities by the Shiba states and is illustrated
 in Fig. \ref{fig02p} (f). As we discuss below, the different scenarios depend on the distance $r$ between the impurities within the range $r>\lambda_F$. For smaller distances, the impurities basically behave as a single impurity
 with total momentum $2S$. The critical coupling for the quantum phase transition where the net effective impurity is fully screened is, thus, $J_0/2$.

% \subsubsection{The role of the distance and spin orientations}
 \begin{figure}[t]\begin{center}
  \includegraphics[width=\columnwidth,height=7cm]{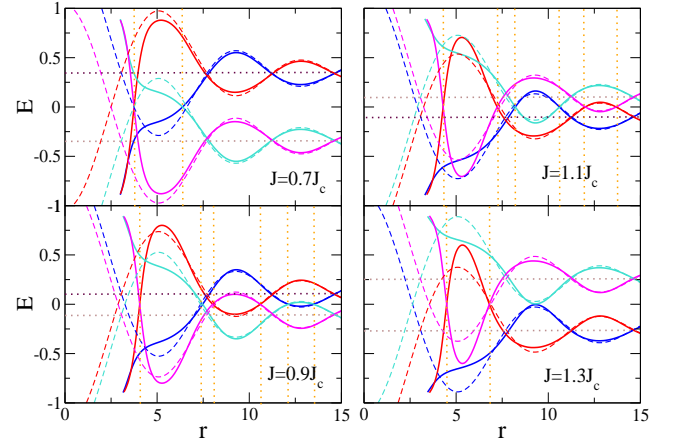}
    \caption{(Color online) Spectrum of Shiba states for a dimer with parallel orientation of the spins along the $z$ axis, within the sector $x^2-y^2$ for different strengths of the exchange interaction $J$ (in units of $J_c$) as a function of the distance between the impurities $r$ (in units of $k_F^{-1}$).  
     $J_c= 1/(\pi \nu S)$ is  the critical coupling for the phase transition of a single impurity. The coherence length of the superconductor is $\xi_0= 100/k_F$. The exact solution is plotted in solid lines, while dashed lines correspond to the spectrum 
   calculated with the effective Hamiltonian. Dark (light) plots correspond to states polarized antiparallel (parallel) to the impurity. 
   Blue and red correspond to $+$ and $-$ combinations. Dark and light dotted horizontal lines indicate  the energies $E_{x^2-y^2}^{\pm}$
   of the single impurities.
   The vertical lines indicate phase transitions.}
 \label{fig1}
\end{center}\end{figure}
 
 We show now some results for the spectrum of a YSR dimer calculated by exactly calculating the subgap Green's function, as explained in Section \ref{sec:formal}, as well as from the solution of the effective Hamiltonian. 
  In order to analyze the effect of the distance, we start by focusing on the channel $x^2-y^2$. 
  Results for  parallel spins oriented along the $z$-direction  are shown in Fig. \ref{fig1}. 
  The Fig. shows the spectrum as a function of the distance between impurities
 for four different values of $J$.   
 It is convenient  to notice that  the configuration where the two impurities are far apart,  the physics is expected to be similar to that of a single impurity.  
 In the long-distance behavior of the dimer we can identify the asymptotic limit to the energies of the YSR states of a single impurity, $E_{x^2-y^2}^{\pm}$ given by Eq. (\ref{emu}). The latter are indicated in dotted lines in the Fig. 
 For parallel spins, the interaction mediated by the substrate is only a hopping term in the framework of the effective Hamiltonian of Eq. (\ref{hefjjp}),
 \begin{equation}
 H_{\rm eff, parallel} = \epsilon_0 \left( c^{\dagger}_1 c_1 + c^{\dagger}_2 c_2 \right) + t  \left(c^{\dagger}_1 c_2 + h.c \right).
 \end{equation}
 This can be easily verified by evaluating the scalar products entering Eq. (\ref{heod}). Expressed in the operators ${\bf c}^{\dagger}=(c^{\dagger}_1,c^{\dagger}_2, c_1,c_2)$, the effective Hamiltonian matrix reads
 ${\cal H}_{\rm eff, parallel}= \epsilon_0 \tau_z + t s_x \tau_z$, where $\tau_z$ is the $z$  Pauli matrix defined in the particle-hole degrees of freedom, while $s_x$ is the $x$ Pauli matrix for the impurity indices $1,\;2$.
 The eigenenergies are $E_{\pm, +}=  |\epsilon_0 \pm t |$ and $E_{\pm, -}= - |\epsilon_0 \pm t |$.   Here, the first index $\pm$ refers to bonding (+) and antibonding (-) combinations of the single impurity states, while the second corresponds
 to ${\rm s}=\pm$ as in Section \ref{sec:formal}.

 Hence, the hybridization leads to  the formation of  bonding  and antibonding combinations, which are polarized antiparallel or parallel to the impurities. Focusing on the long-distance region of the plots and comparing the different panels, we can identify the transition to the bound state of the 
 single impurity, as $J$ increases and overcomes the critical value $J_c$. Notice that  states plotted in dark and light colors cross as $J > J_c$. We see that the agreement between the exact solution and the prediction of the effective Hamiltonian is excellent for distances $r>\lambda_F$. When the two impurities are very close ($r<\lambda_F$), only the exact solution is reliable. The physical picture of this regime can be understood in terms of a single impurity
 with an effective spin $S^{\rm eff}= 2 S$, which corresponds to $J_c^{\rm eff}=J_c/2$, as mentioned before. For this reason, even when the Shiba states of the single isolated impurity are not bound, they become bounded when the impurities are very close one another. For intermediate distances, many transitions may take place, depending on the value of the exchange interaction, which are indicated by vertical lines in the Fig. For the case of $J=0.7J_c$, shown in the upper left panel of the Fig., 
 there are two phase transitions. Starting from the large-$r$ limit, there is a transition to a phase where the  $+$ state is bound, while the $-$  remains unbounded. As $r$ decreases further, there is a transition to the short-distance phase, where both
 molecular states ($+$ and $-$)  are bound. For $J=0.9J_c$, shown in  the left bottom panel of the Fig., several phase transitions take place. Starting from the two unbound states in the long-distance regime, the $+$ and 
 $-$ states alternate to get bound
 and unbound as $r$ decrease. We can observe a similar situation for $J=1.1 J_c$ (upper right panel), but in this case  the long-distance phase corresponds to the two molecular states bound. For even larger $J$, as in the case shown in the bottom 
 right panel, the number of intermediate transitions decrease, and there is only a single phase with only one molecular bounded state. In all the cases shown in the figure, the short-distance regime ($r<\lambda_F$) correspond to two states bounded to 
 an effective impurity with $S^{\rm eff}$. Following Ref. \onlinecite{moca}, we can make contact to the Kondo  effect taking place in a dimer of quantum spins. The regime with two bound states corresponds to maximum screening while the case of a single bound state corresponds to an orbital doublet and underscreened Kondo effect.

  \begin{figure}[t]\begin{center}
  \includegraphics[width=\columnwidth,height=6.7cm]{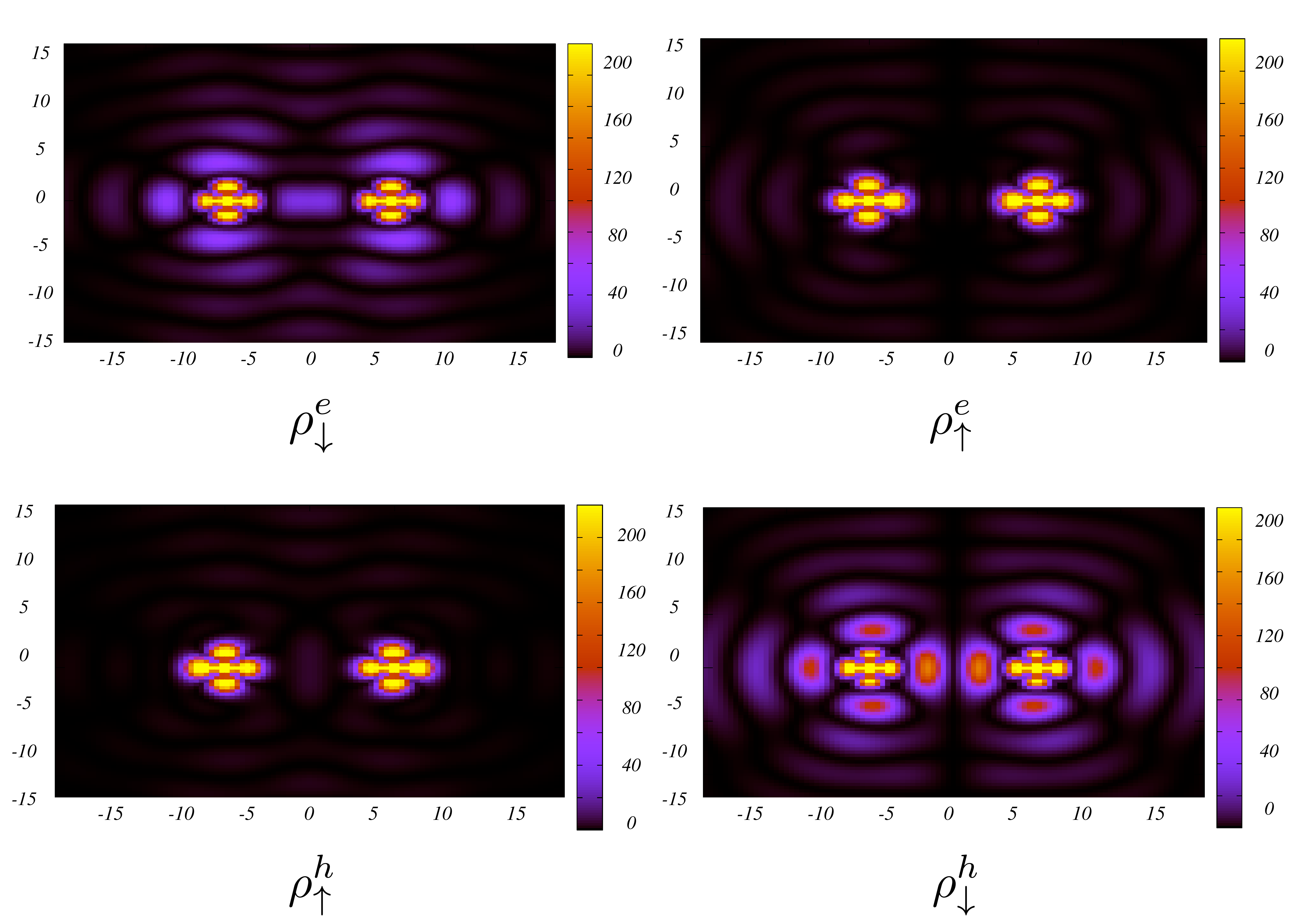}
  \caption{(Color online) Spectral function $\rho^{e,h}_{\sigma}(x,y, 0)$ for the Shiba states  within the channel $x^2-y^2$  for impurities with parallel spins separated in $r=10$, and exchange interaction $J=0.9 J_c$. Both states
  correspond to the positive-energy region of the spectrum.
  The left (right) panels correspond to the state with highest  (lowest) energy, which are bonding (antibonding) configurations.
     Other details are the same as in Fig. \ref{fig1}.}
 \label{fig1p}
\end{center}
\end{figure}

We now show the  spectral densities defined in Eq. (\ref{rho2}) of the YSR states of a dimer, for the particular cases of the spectra shown in the previous section. 
We have verified that the spectral density calculated with the eigenstates of the effective Hamiltonian, as defined in Eq. (\ref{rhoeff}), reproduces the exact result for the distance between impurities shown in Fig. \ref{fig1p}.
The latter can be computed from the eigenstates of the effective Hamiltonian and the transformation of Eq.  (\ref{states}) and  (\ref{trans})  with $\theta_1=\theta_2=\phi_1=\phi_2=0$.
 Fig. \ref{fig1p} corresponds to the dimer with parallel spins in the ${x^2-y^2}$ channel
shown in Fig. \ref{fig1}.
 The spectral density provides information on the
nature of the excited states. In fact, the spacial distribution enables the identification of a bonding or antibonding combination. In addition, as in the case of a single impurity, excitations with positive energy and dominant particle component with spin antiparallel to the impurity imply a ground state without bound states in that channel for that configuration. Instead, excitations with positive energy and dominant hole component with spin aligned with
the impurity implies a ground state with a bound state in that configuration and channel. In the example of Fig. \ref{fig1p} we can identify the bonding (antibonding) configuration in the left (right) panels. The analysis of the
 components of the density of states resolved in spin and charge conjugation reveals that the ground state in the first case is unbounded, while it is bounded in the second one, in agreement with the analysis of the spectrum
 shown in Fig. \ref{fig1p}. 
 The substrate and the crystal field splitting strongly affects the shape of the maps for the spectral densities. As an example, we can compare with Fig. \ref{figilu}, which  corresponds to the the bonding state with positive energy  for the same parameters as Fig. \ref{fig1p}.
 The only difference between the configurations of Figs. \ref{figilu} and \ref{fig1p} is the orientation of the line connecting the two impurities relative to the axis of the crystal field. We see that the pattern of the spectral density differs. However, in both cases we can identify a bonding configuration.

  \begin{figure}[t]\begin{center}
  \includegraphics[width=\columnwidth,height=6.7cm]{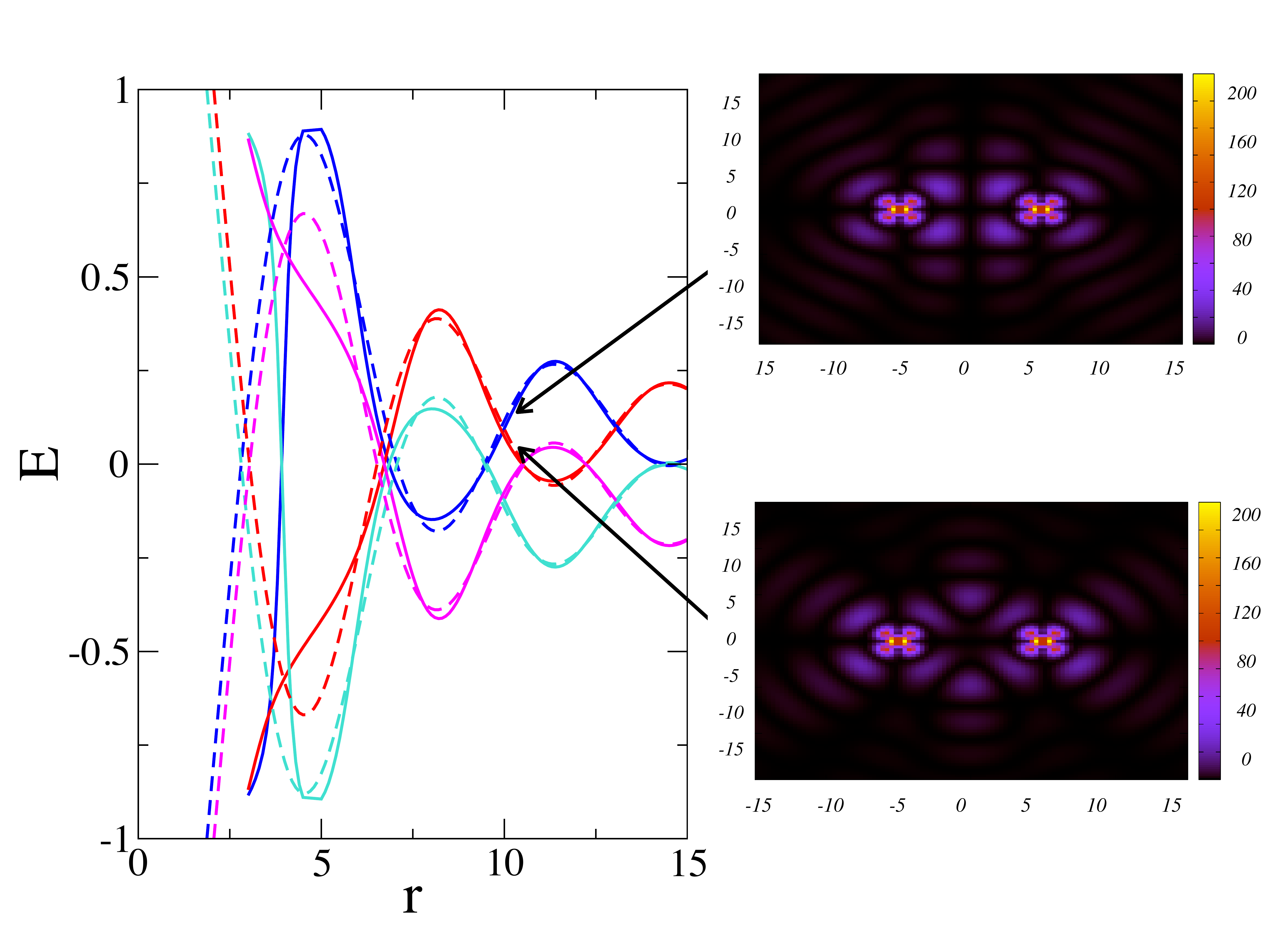}
  \caption{(Color online) Spectrum as a function of the distance and spectral function $\rho^{e}_{\sigma}(x,y, 0)$ for selected Shiba states within the channel $xy$ for impurities with parallel spins separated in $r=10$. The hole components, $\rho^{e}_{\uparrow}(x,y, 0)$, are fainter and they are not shown. The exchange interaction is $J=0.9 J_c$. 
     Other details are the same as in previous Figs.}
 \label{figxy}
\end{center}
\end{figure}

\begin{figure}[t]\begin{center}
  \includegraphics[width=\columnwidth,height=7cm]{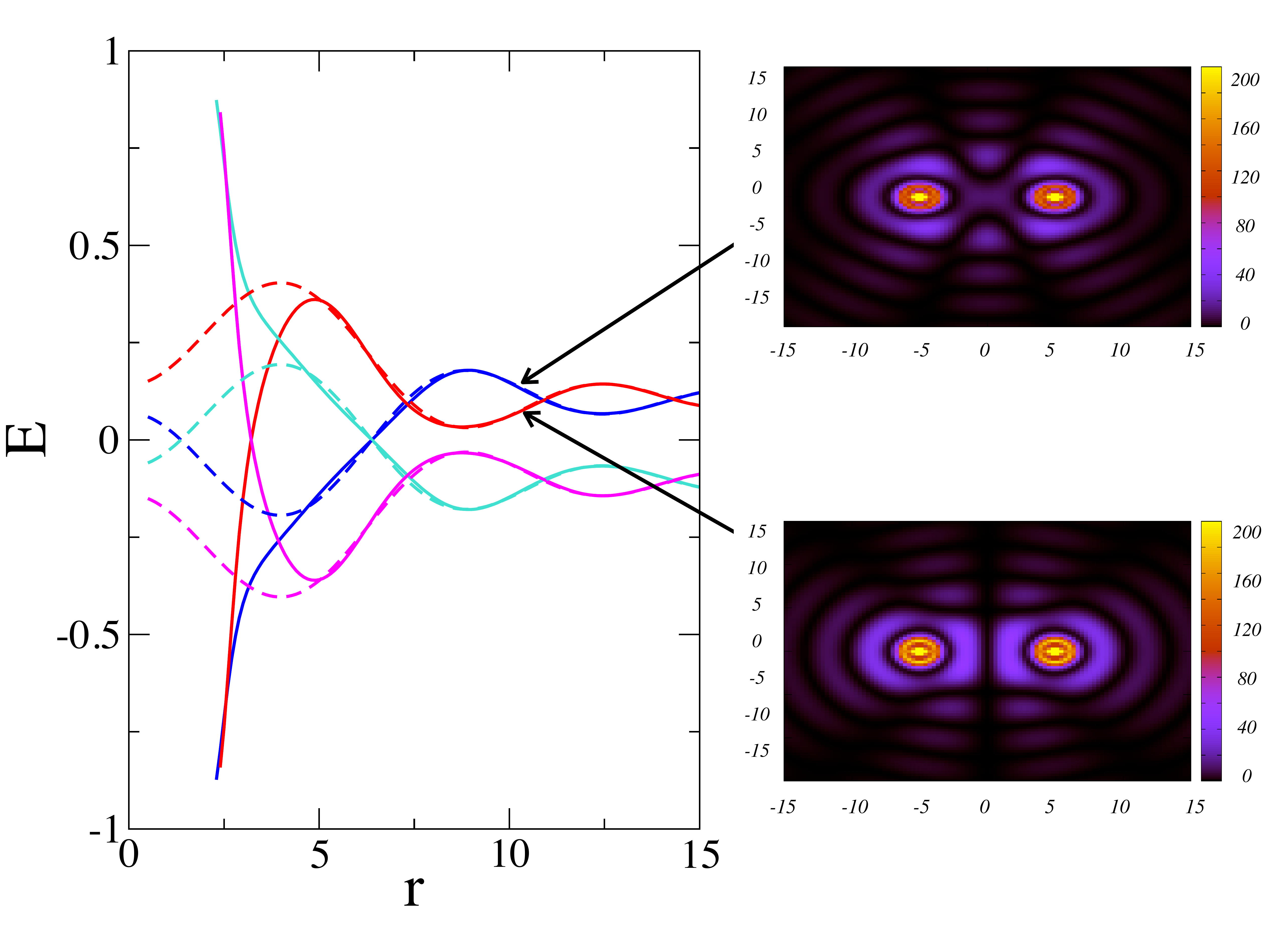}
  \caption{(Color online) Spectrum as a function of the distance and spectral function $\rho^{e}_{\uparrow}(x,y, 0)$ for selected Shiba states within the channel $z^2$ for impurities with parallel spins separated in $r=10$. The exchange interaction is $J=0.9 J_c$. The hole components, $\rho^{h}_{\uparrow}(x,y, 0)$, are fainter and they are not shown. 
     Other details are the same as in previous Figs.}
 \label{figz2}
\end{center}
\end{figure}

\begin{figure}[t]\begin{center}
  \includegraphics[width=\columnwidth,height=7cm]{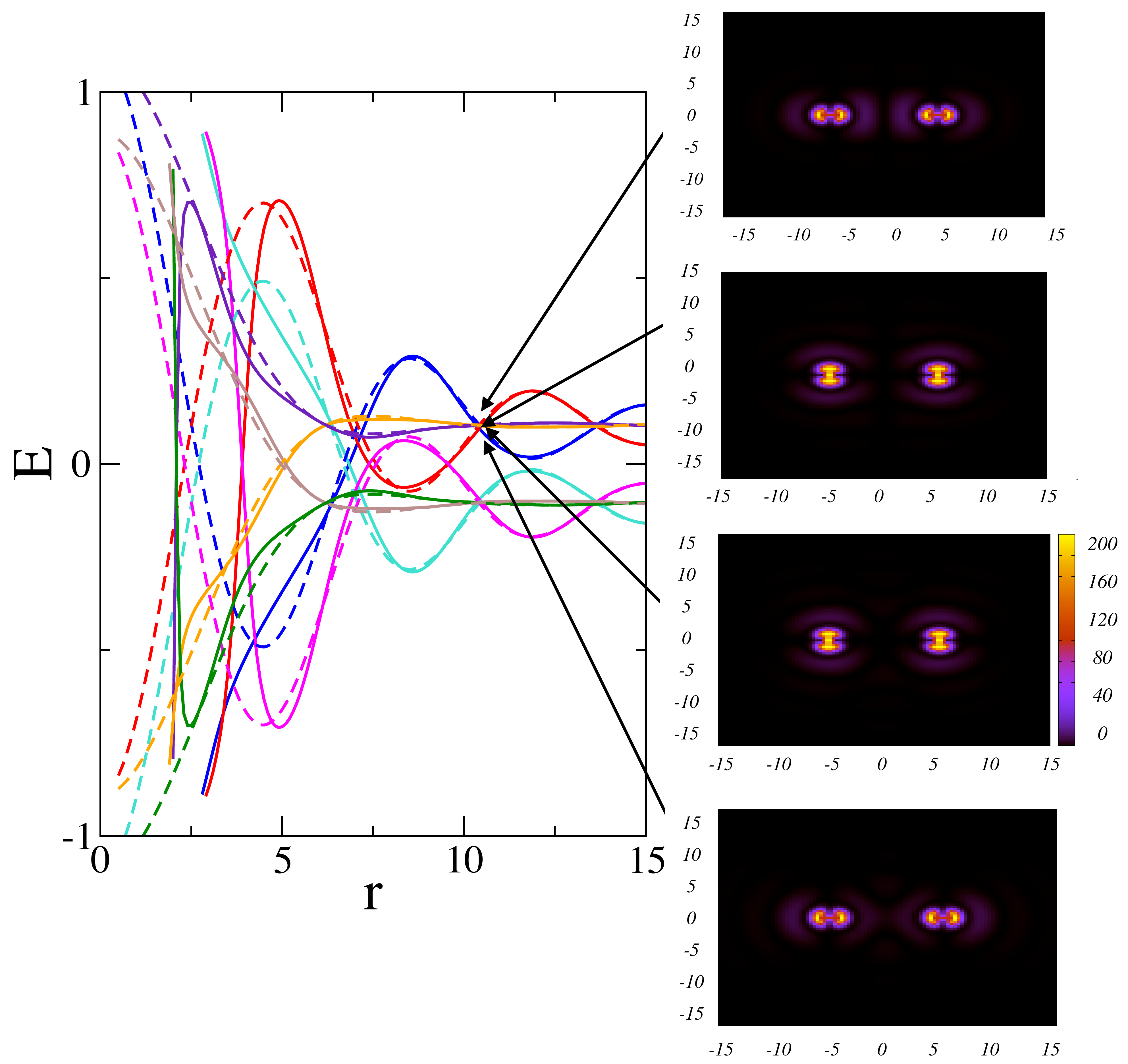}
  \caption{(Color online) Spectrum as a function of the distance and spectral function $\rho^e_{\uparrow}(x,y, 0)$ for selected Shiba states within the degenerate channels $xz,\;yz$ for impurities with parallel spins separated in $r=10$.  The exchange interaction is $J=0.9 J_c$.      Other details are the same as in previous Figs.}
 \label{figxyz1}
\end{center}
\end{figure}

\begin{figure}[t]\begin{center}
  \includegraphics[width=\columnwidth,height=7cm]{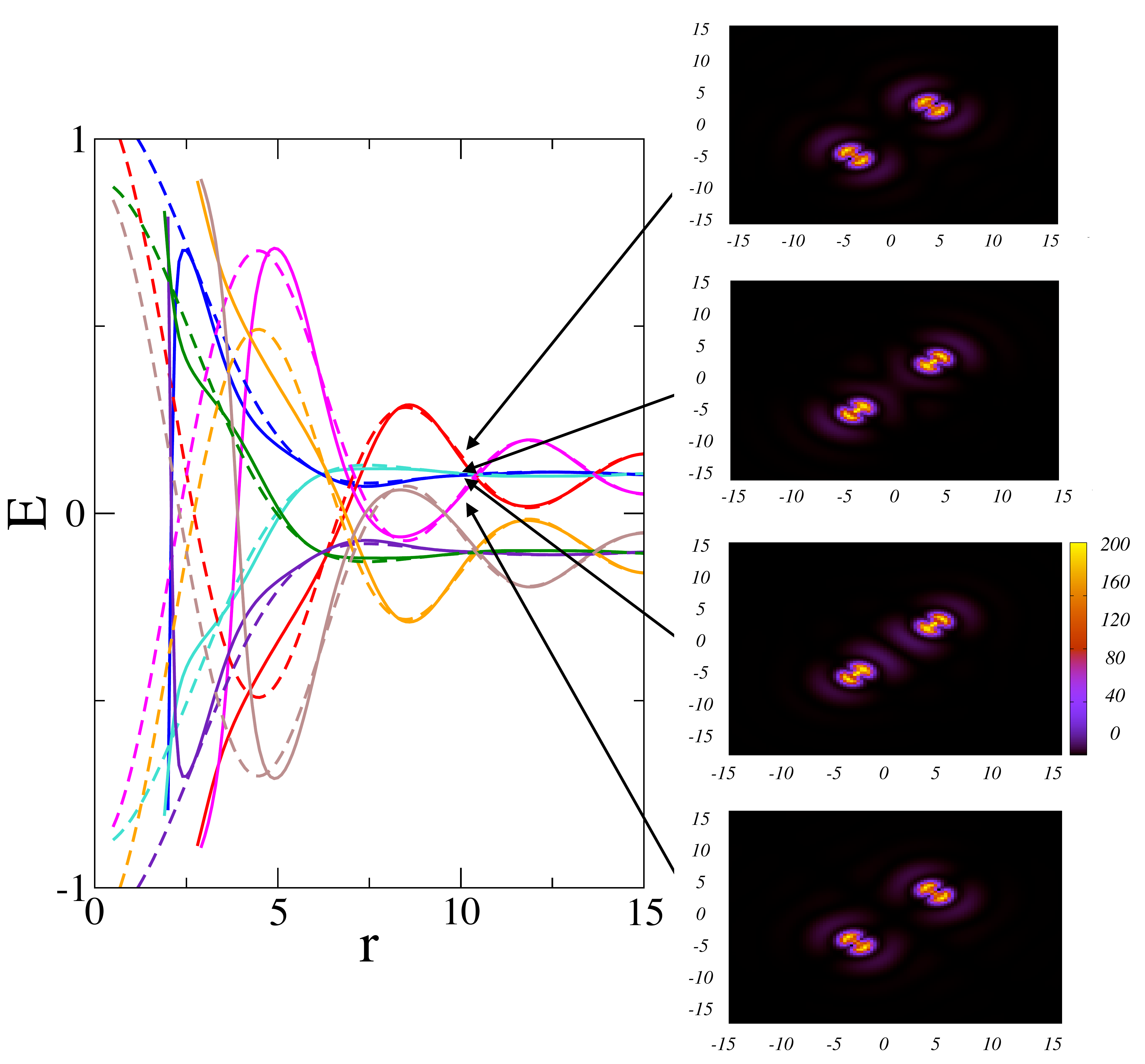}
  \caption{(Color online) Spectrum as a function of the distance and spectral function $\rho^{e}_{\uparrow}(x,y, 0)$ for selected Shiba states within the degenerate channels $xz,\;yz$ for impurities with parallel spins separated in $r=10$.  The exchange interaction is $J=0.9 J_c$. The dimer is oriented forming an angle $\pi/4$  with respect to the $x$ axis of the crystal field. 
     Other details are the same as in previous Figs.}
 \label{figxyz2}
\end{center}
\end{figure}

The behavior of the spectrum and the spectral densities for the other channels is illustrated in Figs. \ref{figxy}, \ref{figz2}, \ref{figxyz1} and \ref{figxyz2}. We can see that, although the asymptotic states in the limit of large distance between
impurities are the same for all the channels, as a function of the distances  there are a large number of crossings and the ordering of the states changes significantly. For the selected cases where the spectral densities  are shown,
we see that we can easily identify the bonding and antibonding configuration in the Figs.  \ref{figxy} and \ref{figz2}. In the case of Figs. \ref{figxyz1} and \ref{figxyz2} the spectral densities are fainter for the value $z=0$ chosen
for the plots, as a consequence of the small projection of the $xz$ and $yz$ orbitals on this plane. In the cases shown, all the ground states are unbound, which is reflected by the dominant spectral density in the particle sector with spin antiparallel to the impurities. The results shown in Figs. \ref{figxy} and \ref{figz2}
correspond to the dimer oriented in different directions with respect to the axis of the crystal field. Comparing the two figures, we see that the orientation does not have a major impact in the spectrum. 
However, it plays a role in the spacial distribution of the wave functions. In the orientation shown in Fig. \ref{figxyz2} the two orbitals hybridize to form configurations of the type $d_{xz} \pm d_{yz}$ and the spectral density modifies accordingly. These details are a consequence of the type of substrate and  the crystal field.

\subsection{Dimer with tilted spins}

 \begin{figure}[t]\begin{center}
  \includegraphics[width=\columnwidth,height=7cm]{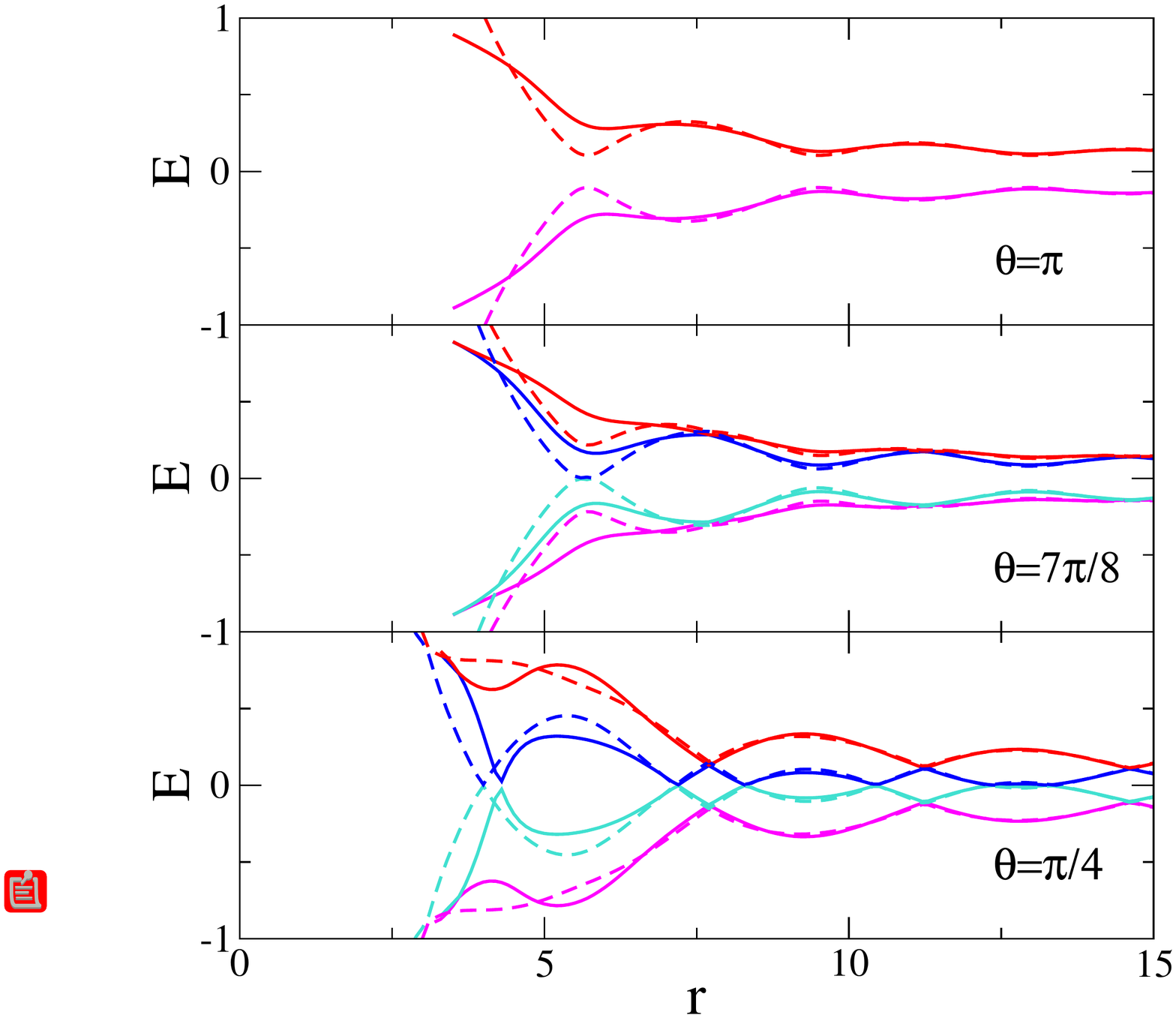}
    \caption{(Color online) Spectrum of Shiba states for a dimer with tilted magnetic moments $\theta_1=0, \theta_2=\theta$ with respect to the $z$ axis, and  $J=0.9 J_c$, as a function of the distance between the impurities $r$.    
    Other details are the same as in Fig. \ref{fig1}.}
 \label{figap}
\end{center}\end{figure}

When the spins of the impurities are tilted by an angle $\theta= \theta_1-\theta_2$, the midgap states not only hybridize but also become paired with an effective $p$-wave  interaction. Hence, the ground state of the full system consists of the two unscreened impurities
in the BCS condensate of the substrate plus the BCS state of the paired subgap states. This is illustrated in the sketches of Figs. \ref{fig02a} and \ref{fig03}. 
The Shiba states in this case correspond to excitations of the BCS state with $p$-wave pairing. 
\begin{equation} \label{hef12}
 H_{\rm eff, \theta} = \epsilon_0 \left( c^{\dagger}_1 c_1 + c^{\dagger}_2 c_2 \right) +t  \; c^{\dagger}_1 c_2 +   \Delta c^{\dagger}_1 c^{\dagger}_2 + h.c.
 \end{equation}
 In  Nambu language, this can be represented by the matrix ${\cal H}^{eff}_{\theta}= \epsilon_0 \tau_z + t s_x \tau_z+ \Delta s_y \tau_y $.
 The corresponding 
eigenenergies are 
\begin{equation}
E_{\pm, \pm}=\pm \sqrt{\epsilon_0^2 + t^2 +\Delta^2 \pm 2 t \sqrt{\Delta^2+ \epsilon_0^2}},
\end{equation}
where, as before, the second label ${\rm s}=\pm$ indicates positive and negative energies and the first  $\pm$ labels the two possible solutions in each case.
 The spectrum is always gapped for $|t| < \sqrt{\Delta^2 + \epsilon_0^2}$. This is the case, in particular, for impurities with antiparallel magnetic moments ($\theta=\pi$), where It can be verified that the hopping term  of Eq. (\ref{heod}) vanishes ($t=0$), while the pairing term is finite.   
   
Example of the spectra for dimers of impurities with a relative angle $\theta$ between the magnetic moments are shown 
 in Fig. \ref{figap} within the channel $d_{x^2-y^2}$. The upper panel corresponds to the antiparallel configuration. 
 As before it is useful to start the analysis in the long-distance regime, where the inter-impurity effects are very small and
we expect that the YSR states correspond to states that are basically the corresponding ones of the single impurity.
 As the impurities become closer, an hybridization takes place between $\downarrow$ quasiparticles localized at the impurity $1$ and
$\uparrow$ quasiholes localized at the impurity $2$. The spectrum is always gapped in this case. Since the total spin $S_z=0$, each of these excited states have a doubly degeneracy. 
While the behavior of the dimer of parallel spins can be related to the Kondo effect of quantum spins, this is not the case of the antiparallel configuration. 
Other angles of the relative orientation of the magnetic moments of the dimer are shown in the other panels of the Fig. \ref{figap}. The lower panel corresponds to a tilt closer to the parallel configuration of the impurities. The amplitude of the hopping, $|t|$, increases as $\theta$ decreases and  is a function of the distance between the impurities.
For some distances, the hopping overcomes the critical value
$t_c=\sqrt{\Delta^2 + \epsilon_0^2}$, for which the gap closes and level crossings at zero energy are observed in the spectrum. As before, this can be interpreted as a quantum phase transition where the parity of the ground state changes.
The nature of the bound state is different from that of the fully polarized system. In fact, impurities with parallel magnetic moments the excitations and, in particular, the bound states, are simple bonding and antibonding combinations of the 
YSR states localized at the individual impurities. Instead, in the tilted case, the latter form a p-wave BCS state, which is degenerate with that of the substrate and the subgap excitations are the excitations of this state of p-wave pairs. 
When the dimer has a net polarization, these excitations disperse in bonding and antibonding combinations and eventually some of them can cross zero energy and get bounded to the net magnetic moment of the dimer.

% \begin{figure}[t]\begin{center}
 % \includegraphics[width=\columnwidth,height=7cm]{dx2-y2-ap-2.pdf}
  %  \caption{(Color online) Spectral function $\rho^{e,h}_{\sigma}(x,y, 0)$ for the two degenerate Shiba states with positive energy within the channels $x^2-y^2$ for impurities with antiparallel orientation of the spins along the $z$ separated in $r=10$.  The exchange interaction is $J=0.9 J_c$.   The contributions of the two degenerate states have been added.   Other details are the same as in previous Figs.}
% \label{figap-rho}
%\end{center}\end{figure}

%The spectral densities $\rho^{e,h}_{\sigma}(x,y, 0)$ are shown in Fig. \ref{figap-rho} for  the states with positive energy. Since there are two degenerate states, the plots show the total spectral density that results from adding the contributions
%of the two degenerate states. Unlike the case of parallel impurities, we see now components of the spectral function in the two spin components, as well as in the particle and hole degrees of freedom. 

 \begin{figure}[t]\begin{center}
  \includegraphics[width=\columnwidth,height=6.5cm]{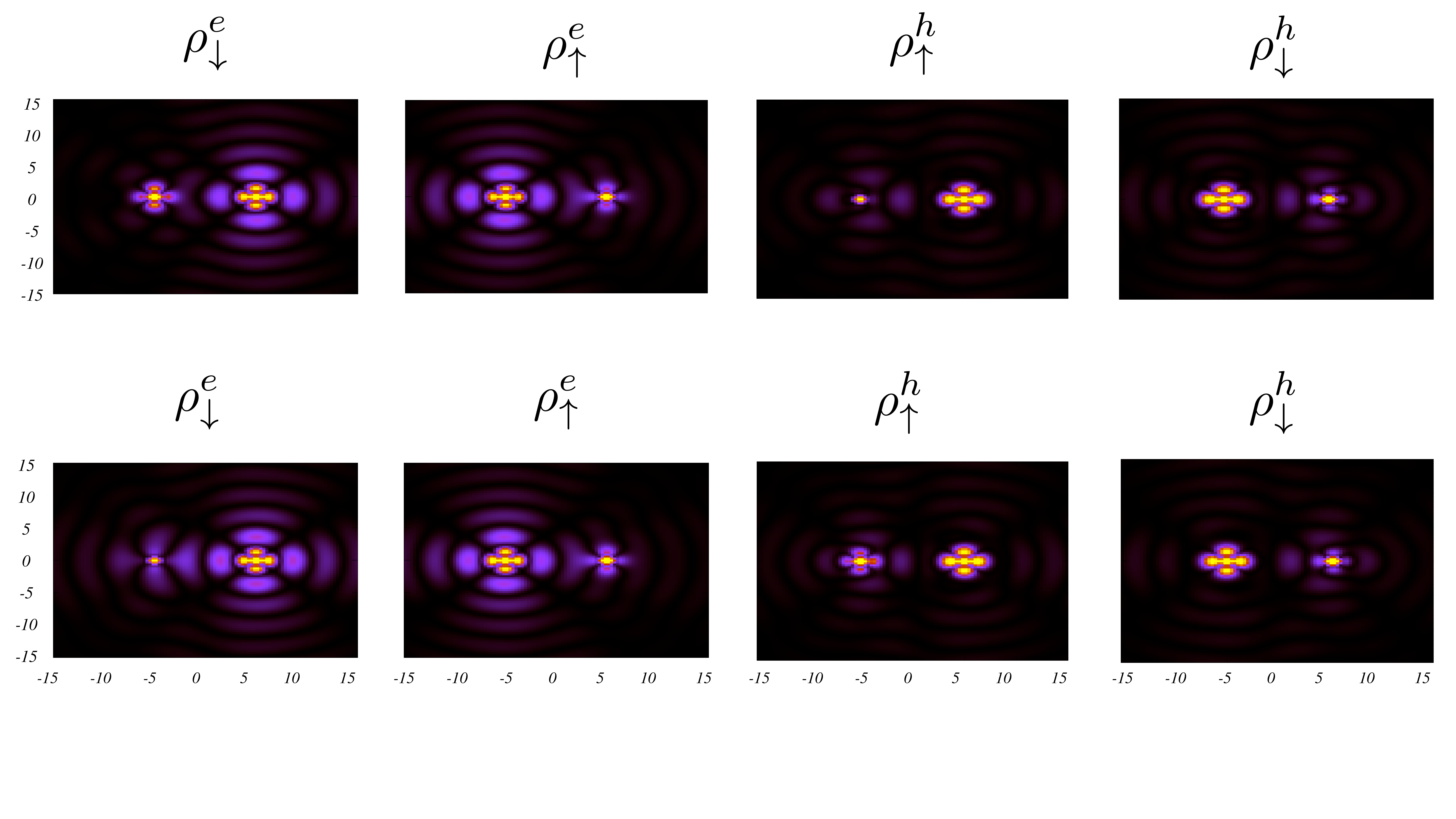}
    \caption{(Color online) Spectral function $\rho^{e,h}_{\sigma}(x,y, 0)$ for the two Shiba states with positive energy within the channels $x^2-y^2$ for impurities with a tilt of $\theta=7\pi/8$ in the magnetic moments separated in $r=10$.  Upper and lower panels corresond, respectively to highest and lowest energies. The exchange interaction is $J=0.9 J_c$.      Other details are the same as in previous figures.}
 \label{fig-rho-tilt}
\end{center}\end{figure}

We have shown results for a single channel, but similar results are obtained in the different channels and only the shape of the wave-function configuration  change.
In these figures, we have shown results obtained by exactly calculating the subgap Green's function. In all the cases shown, we have compared with the spectral densities calculated with the effective Hamiltonian and we have verified
that the  agreement is excellent.

\begin{figure}[t]\begin{center}
  \includegraphics[width=\columnwidth,height=6.5cm]{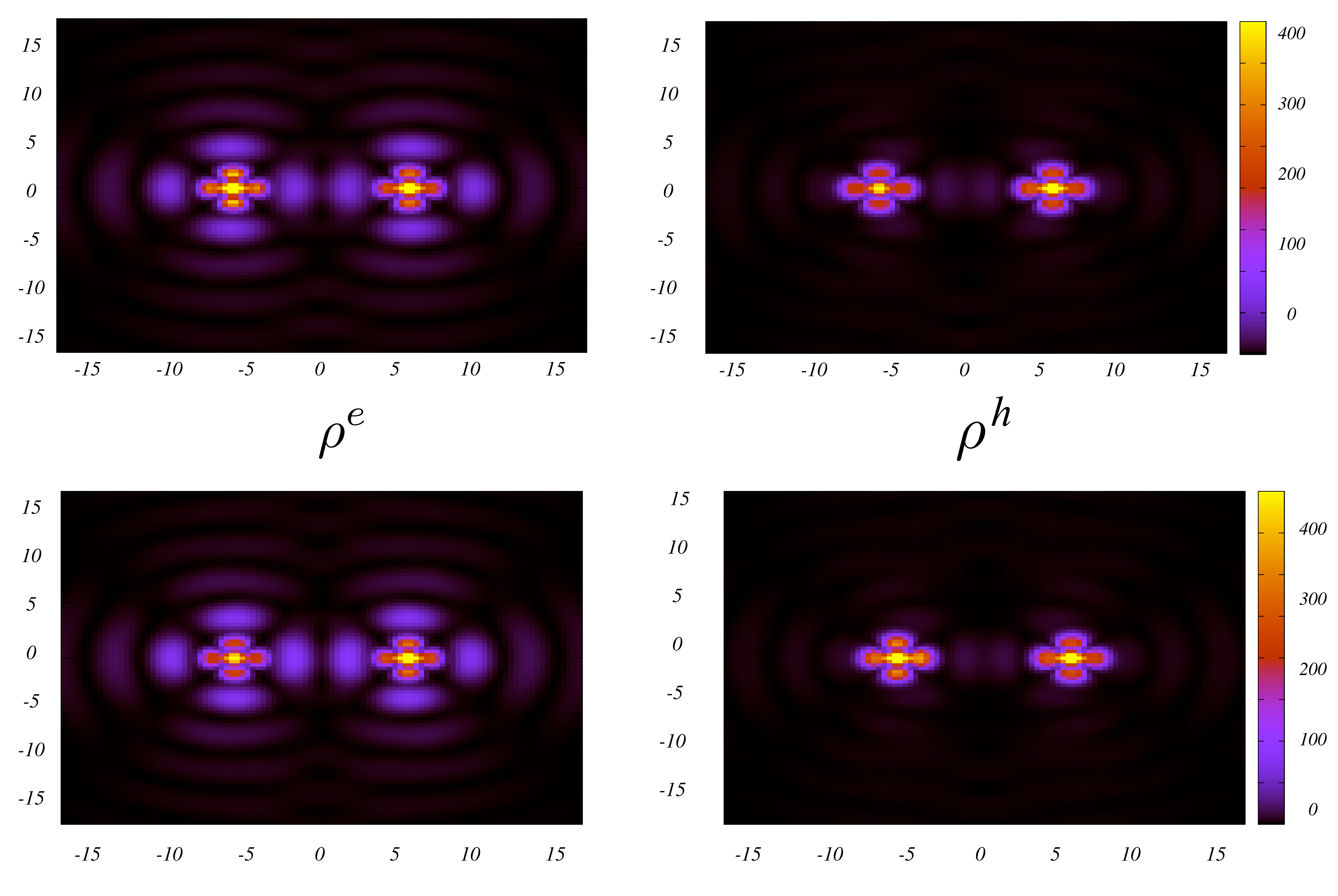}
    \caption{(Color online) Spectral function $\rho^{e,h}(x,y, 0)=\sum_{\sigma}\rho^{e,h}_{\sigma}(x,y, 0)$ for the two Shiba states with positive energy within the channels $x^2-y^2$ for impurities with a tilt of $\theta=7\pi/8$ in the magnetic moments separated in $r=10$.      Other details are the same as in the previous figure.}
 \label{fig-rhot-tilt}
\end{center}\end{figure}

In Fig. \ref{fig-rho-tilt} we illustrate the behavior of the spectral density of a dimer with magnetic moments forming an angle $\theta=7 \pi/8$ in the orbital channel $d_{x^2-y^2}$. The total spectral density corresponding tho this configuration
 is shown in Fig. \ref{fig-rhot-tilt}.

\begin{figure}[t]\begin{center}
  \includegraphics[width=\columnwidth,height=6.5cm]{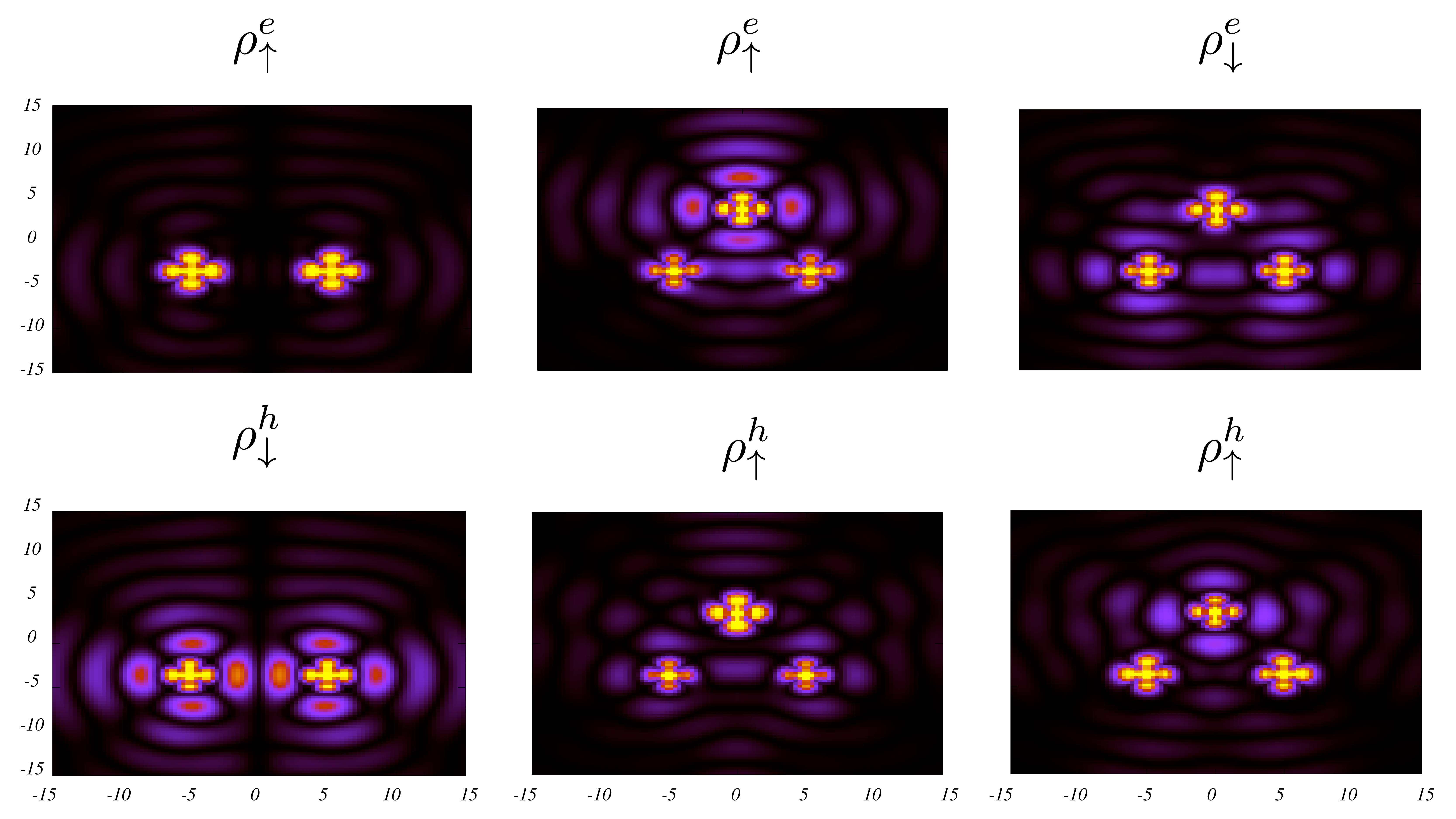}
    \caption{(Color online) Spectral function $\rho^{e,h}_{\sigma}(x,y, 0)$ for the degenerate Shiba states with positive energy within the channels $x^2-y^2$ two impurities ferromagnetically aligned along the $z$ axis and a third impurity with antiparallel orientation with respect to the other two. The three impurities are separated by the same distance $r=10$, forming 
    an equilateral triangle.  The exchange interaction is $J=0.9 J_c$.   Left (right) configurations correspond to  the state with lowest  (highest) energy.  The vanishing components are not shown.   Other details are the same as in the previous figure.}
 \label{tri-frus}
\end{center}\end{figure}

\subsubsection{Trimer with magnetic frustration}
We close with an illustration of the trimer configuration. 
In the case of three impurities, there are several possible scenarios, which depend on the orientations of the magnetic moments as well as on the spacial configuration of the cluster. We are not going to do an exhaustive
analysis of the trimer, but just illustrate here how the previous techniques and analysis can be extended to study more complex clusters. 

When the three impurities are  ferromagnetically alined,
the expected scenario is basically a generalization of the one discussed for the ferromagnetic dimer. The antiferromagnetic configuration  is, however, frustrated in the present case. In what follows, we focus on such situation, which corresponds 
to two of the impurities with the magnetic moments ferromagnetically aligned and the third one antiferromagnetically oriented with respect to the other two. 
The effective Hamiltonian for impurities separated by $r> \lambda_F$ is
 \begin{equation}\label{trimer}
 H_{\rm eff}^{\uparrow \uparrow \downarrow} = \epsilon_0 \sum_{j=1}^3  c^{\dagger}_j c_j  +  t_{12} c^{\dagger}_1 c_2  + \Delta_{23} c^{\dagger}_2 c^{\dagger}_3 +
 \Delta_{31} c^{\dagger}_3 c^{\dagger}_1+ H.c. ,
 \end{equation}
which corresponds to a tight-binding hopping between the states localized at the two parallel impurities (labeled with $j=1,2$)  and a p-wave pairing  term between the latter and the states localized at the third one (labeled with $j=3$). 
For the case where the cluster is spatially organized forming an equilateral triangle with the parallel impurities placed on the $x$ axis and the antiparallel one on the $y$-axis, we can see that $\Delta_{23}=-\Delta_{31}=\Delta$. Then
we can rewrite the effective Hamiltonian of  Eq. (\ref{trimer}), in terms of the bonding and antibonding combinations of states localized at $1$ and $2$, respectively, $c_{\pm} = \left( c_1 \pm c_2 \right)/\sqrt{2}$. The result is
 \begin{equation}\label{trimer1}
 H_{\rm eff}^{\uparrow \uparrow \downarrow} = \epsilon_0 c^{\dagger}_3 c_3+ \sum_{j=\pm}  E_j c^{\dagger}_j c_j  +  \sqrt{2}{\Delta}  \left( c^{\dagger}_+ c^{\dagger}_3 + H.c. \right),
 \end{equation}
 which corresponds to a pairing interaction between the third impurity and the bonding configuration of 1 and 2, while  the antibonding configuration only couples the parallel impurities $j=1,2$. This implies that only the antibonding configuration
 can bound to screen the parallel impurities while the bonding will always form an effective BCS state with the localized states of the third impurity. The corresponding density of states for these excitations  are shown in Fig. \ref{tri-frus} focusing on the
 $d_{x^2-y^2}$ channel. For the
 parameters chosen, the lowest-energy state with positive energy has the spectral density shown in the left panels of Fig. \ref{tri-frus} and corresponds to an excitation where the ground state is formed with the antibonding state
 bound to the parallel impurities. The other two states are the two excitations of the paired bonding state and the localized state at the third impurity. Since the total spin of the impurities is different from zero in this case,
 the excitations with $\uparrow$ and $\downarrow$ spins are not degenerate and their spectral densities are shown in the central and right panel of Fig. \ref{tri-frus}.

\section{Summary and conclusions}\label{sec:sum-con}
We have presented a systematic theoretical framework to analyze the spectrum and density of states of Yu-Shiba-Rusinov states of clusters of magnetic adatoms with multiple active orbitals and large magnetic moments in superconducting substrates.
The treatment is based on the formulation of a multiorbital Kondo Hamiltonian considering the total spin of the impurity as a classical magnetic moment. We have defined an effective Green's function to describe the subgap states, such that the poles of this function coincide with the poles of the $T$-matrix of the problem. Given this Green's function, we can calculate the contribution of the subgap states with the local density of states and also
define an effective Hamiltonian to describe clusters of diluted impurities. 

We have analyzed the case of dimers in substrates with constant density of states assuming the effect of the crystal field in the Kondo model. We have analyzed different relative orientations of the magnetic moments and found an excellent agreement between the exact solution of the problem and the description based on the effective Hamiltonian. We have successfully described not only the spectrum but also the local density of states in the different orbital channels. We have briefly analyzed the case of a trimer.

For simplicity we focused on a simple substrate with s-wave superconductivity modeled by a BCS Hamiltonian expressed in the basis of plane waves. The  present systematic approach to evaluate the subgap Green's function can be also
implemented in models for more realistic substrates, which include lattice effects, several bands and spin-orbit interaction. This 
 method can be also useful to go beyond the  BCS Hamiltonian model with a rigid gap and to calculate the correction to the local gap due to the 
presence of the impurity. Such correction was found to be relevant, in particular, for impurities localized at energies close to the onset of he quasiparticle continuum. \cite{meng1,meng2}
In the present framework, the first correction to the bare gap due to the impurity can be calculated by solving the gap equation with the anomalous component of the Green's function of Eq. (\ref{greenr}), 
and the T-matrix evaluated by recourse to the subgap Green's function as expressed in Eq. (\ref{tg}). We have verified that the latter approximation leads to an accurate description of the subgap spectrum and the density of states, not only for Shiba states deep in the gap but also in cases where their energies are close to the gap, hence it is reasonable to expect that it can be also useful to compute effects related to the gap renormalization. 

\section{Acknowledgements}
The author thanks Felix von Oppen for interesting discussions,  as well as A. A. Aligia and N. Lorente for stimulating comments. The author also thanks the hospitality of the Dahlem Center for Complex quantum Systems, Berlin under the support of the Alexander von Humboldt Stiftung, Germany. LA also acknowledges 
support from PIP-2015-CONICET, PICT-2017, PICT-2018, Argentina, and
Simons-ICTP-Trieste associateship Argentina.

\begin{widetext}

\appendix
\section{Calculation of $F$}\label{f}
\subsection{Single impurity}
We start from the definition of the function in Eq. (\ref{def-f}) and consider a constant density of states $\nu \simeq \sum_{\bf k} \delta(\xi-\xi_{\bf k})$. Then,
\begin{equation}
F(\omega) =  \sum_{\bf k}G^0(\xi_{\bf k},\omega)= \int d\xi  \sum_{\bf k} \delta(\xi-\xi_{\bf k}) G^0(\xi,\omega) = \nu \int d\xi G^0(\xi,\omega).
\end{equation}
The non-interacting Green function is
\begin{equation}
G^0(\xi,\omega) = \left[(\omega + i \eta) \tau_0 - \xi \tau_z - \Delta \tau_x \right]^{-1}=\frac{1}{(\omega + i \eta)^2 - \xi^2 - \Delta^2}\left[ (\omega + i \eta) \tau_0 + \xi \tau_z + \Delta \tau_x \right].
\end{equation}
Then
\begin{equation}
F(\omega)= - \nu \int d\xi  \frac{1}{(\xi-i \overline{\xi})(\xi+i \overline{\xi})} \left[\omega \tau_0 + \Delta \tau_x \right] = - \frac{\nu \pi}{\sqrt{\Delta^2 -\omega^2}} \sigma_0 \left[\omega \tau_0 + \Delta \tau_x \right],\;\,\;\;\;\;\;\;\;\; |\omega| < \Delta,
\end{equation}
with $\overline{\xi}= \sqrt{\Delta^2 -\omega^2}$.

\subsection{Two impurities}
The diagonal components of the function defined in Eq. (\ref{fjj}) are $F_{\mu,\mu^{\prime}}^{1, 1}(\omega) =F_{\mu,\mu^{\prime}}^{2, 2}(\omega)=
\delta_{\mu,\mu^{\prime}} F(\omega) $. Here, we calculate the off-diagonal components. 
In the case of a strong crystal-field splitting, it is reasonable to assume that the matrix elements of the $F$ matrix are diagonal in the index $\mu$.
 In order to calculate 
the off-diagonal elements of the matrix $F$ we  must project the real-space 
 components of the function $e^{i  {\bf k} \cdot ({\bf r}_1 - {\bf r}_2)}$ in the same  axis as the orbitals. Therefore, the off-diagonal matrix element reads
\begin{eqnarray}
F_{\mu, \mu^{\prime}}^{1, 2}(\omega) 
& = & (4 \pi)^2 L^3 \sum_{l,m} (i)^{l} Y^{m}_{l}(\hat{r}_{12}) \int \frac{dk}{(2\pi)^3} k^2 j_{l} \left(k r_{12} \right)  G^0( k, \omega) \int d\Omega_k  Y^{m}_{l}(\Omega_k) f_{\mu}(\Omega_k)^* f_{\mu^{\prime}}(\Omega_k )
 \\
&= &4 \pi \nu \sum_{l} (i)^{l}  \; {\cal C}_{\mu,\mu^{\prime}} (l) 
  \int d\xi   j_{l} \left(\left(\frac{\xi}{{\rm v}_F} + k_F \right)r_{12}\right)    G^0( \xi, \omega) =
- 4 \pi \nu \sum_{l} (i)^{l}  \; {\cal C}_{\mu,\mu^{\prime}} (l)   \left[ \left( \omega \tau_0 + \Delta \tau_x \right) I^{(l)}_1(r_{12},\omega) + \tau_z I^{(l)}_2(r_{12},\omega) \right],\nonumber
% \int d\Omega_k Y^{m^{\prime}}_{l^{\prime}}(\Omega_k) Y_2^m(\Omega_k)^* Y_2^m(\Omega_k ),
\end{eqnarray}
with $r_{12}= |{\bf r}_1 - {\bf r}_2|$ and using the following definition of the integrals of Bessel functions
\begin{equation} \label{inte}
I^{(l)}_j(r,\omega) = \int d \xi j_l \left( (\frac{\xi}{v_F}+k_F) r \right) \frac{\xi^{j-1}}{\xi^2 + \Delta^2-\omega^2} ,\;\;\;\;\;j=1,2.
\end{equation}
The calculation of these integrals is detailed in Appendix (\ref{integrals}) for the cases $l=0,2,4$. 
Notice that they have the following structure
\begin{equation}\label{ilj}
I^{(l)}_{j} = \lambda_j(r,\omega)  \kappa^{(l)}_{j}(k_F r),
\end{equation}
where the functions $\lambda_j(r,\omega)$ are defined in Eq. (\ref{lambda12}) and the functions $\kappa$ are defined in Eqs. (\ref{kappa0}), (\ref{kappa2}) and (\ref{kappa4}), respectively. 
The coefficients entering the above expression are
\begin{eqnarray}\label{coef}
{\cal C}_{z^2} (l) & = & \left(\delta_{l,0}+\delta_{l.2} + \delta_{l,4} \right)  C_{2,2,l}^{0,0,0} Y^{0}_{l}(\hat{r}_{12}),\nonumber \\
{\cal C}_{x^2-y^2} (l) & = &\frac{1}{2} \left[ \left( C_{2,2,4}^{2,2,-4} Y^{-4}_{4}(\hat{r}_{12}) + C_{2,2,4}^{-2,-2,4} Y^{4}_{4}(\hat{r}_{12}) \right) \delta_{l,4} +2 \left(\delta_{l,0}+\delta_{l.2} + \delta_{l,4} \right) C_{2,2,l}^{-2,2,0} Y^{0}_{l}(\hat{r}_{12}) \right],\nonumber\\
{\cal C}_{xy} (l)& = & \frac{1}{2} \left[ - \left( C_{2,2,4}^{-2,-2,4} Y^{4}_{4}(\hat{r}_{12}) + C_{2,2,4}^{2,2,-4} Y^{-4}_{4}(\hat{r}_{12}) \right) \delta_{l,4} +2  \left(\delta_{l,0}+\delta_{l.2} + \delta_{l,4} \right) C_{2,2,l}^{-2,2,0}Y^{0}_{l}(\hat{r}_{12}) \right],\nonumber\\
{\cal C}_{xz} (l) & = & \frac{1}{2} \left[ -\left( C_{2,2,l}^{-1,-1,2} Y^{2}_{l}(\hat{r}_{12}) + C_{2,2,l}^{1,1,-2} Y^{-2}_{l}(\hat{r}_{12}) \right) \left( \delta_{l.2} + \delta_{l,4} \right) +2  \left(\delta_{l,0}+\delta_{l.2} + \delta_{l,4} \right) 
C_{2,2,l}^{-1,1,0} Y^{0}_{l}(\hat{r}_{12}) \right],\nonumber\\
{\cal C}_{yz} (l) & = & \frac{1}{2} \left[ \left( C_{2,2,l}^{-1,-1,2} Y^{2}_{l}(\hat{r}_{12}) + C_{2,2,l}^{1,1,-2} Y^{-2}_{l}(\hat{r}_{12}) \right) \left( \delta_{l.2} + \delta_{l,4} \right) +2  \left(\delta_{l,0}+\delta_{l.2} + \delta_{l,4} \right)
 C_{2,2,l}^{-1,1,0} Y^{0}_{l}(\hat{r}_{12}) \right],\nonumber\\
 {\cal C}_{xz,yz} (l) & = &- \frac{i}{2} \left[ \left( C_{2,2,l}^{-1,-1,2} Y^{2}_{l}(\hat{r}_{12}) - C_{2,2,l}^{1,1,-2} Y^{-2}_{l}(\hat{r}_{12}) \right) \left( \delta_{l.2} + \delta_{l,4} \right) \right], \nonumber\\
 {\cal C}_{xy,xz} (l) & = &- \frac{i}{2} \left[\left(C_{4,2,2}^{3,-2,-1} Y^{3}_{4}(\hat{r}_{12}) + C_{4,2,2}^{-3,2,1} Y^{-3}_{4}(\hat{r}_{12}) \right) \delta_{l,4}-  \left( Y^{-1}_{l}(\hat{r}_{12})   C_{l,2,2}^{-1,2,-1} 
  +  Y^{1}_{l}(\hat{r}_{12})  C_{l,2,2}^{1,-2,1}  \right)\left( \delta_{l.2} + \delta_{l,4} \right) \right], \nonumber\\
   {\cal C}_{xy,yz} (l) & = &- \frac{1}{2} \left[\left(C_{4,2,2}^{3,-2,-1} Y^{3}_{4}(\hat{r}_{12}) -  C_{4,2,2}^{-3,2,1} Y^{-3}_{4}(\hat{r}_{12}) \right) \delta_{l,4}-   \left(      Y^{-1}_{l}(\hat{r}_{12})  C_{l,2,2}^{1,-2,1}         - Y^{1}_{l}(\hat{r}_{12})   C_{l,2,2}^{-1,2,-1} 
    \right)\left( \delta_{l.2} + \delta_{l,4} \right) \right]
\end{eqnarray}

The latter depend on integrals of products of three spherical harmonics, which can be expressed in terms of the $3j$ Wigner coefficient as follows
\begin{equation}
C_{l_1,l_2,l_3}^{m_1,m_2,m_3}=   \int d\Omega_k Y^{m_1}_{l_1}(\Omega_k) Y_{l_2}^{m_2}(\Omega_k)^* Y_{l_3}^{m_3}(\Omega_k ) Y_{l_3}^{m_3} (\hat{r}_{12})= \sqrt{\frac{(2 l_1 +1) (2 l_2 +1)(2l_3+1)}{4 \pi}} 
 \left(\begin{array}{ccc}
 l_1 l_2 l_3 \\
 m_1 m_2 m_3 \end{array} \right)
 \left(\begin{array}{ccc}
 l_1 l_2 l_3 \\
0 0 0\end{array} \right) Y_{l_3}^{m_3} (\hat{r}_{12}).
\end{equation}
In our case, we have: $l_1=l_2=2$. The sum over $l, m$ is restricted to the terms satisfying  the selection rules of the $3j$ symbol, namely $m_1+m_2+m_3=0$ and $l=0,2,4$.

We present bellow a table with the values of the different coefficients $C_{l_1,l_2,l_3}^{m_1,m_2,m_3}$ entering (\ref{coef})
\begin{center}
 \begin{tabular}{|| c | c | c  ||} 
 \hline
 $C_{2,2,0}^{0,0,0}$ & 
  $C_{2,2,0}^{-1,1,0}$ & 
   $C_{2,2,0}^{-2,2,0}$ 
  \\ [0.5ex] 
 \hline\hline
  $ \sqrt{\frac{1}{4\pi}}  $&
 $ - \sqrt{\frac{1}{4\pi}}  $&
 $\sqrt{\frac{1}{4\pi}} $ 
 \\  [1ex] 
 \hline
\end{tabular}
\end{center}
 
 \begin{center}
 \begin{tabular}{|| c | c | c  | c | c | c ||} 
 \hline
  $C_{2,2,2}^{0,0,0}$ & 
   $C_{2,2,2}^{-2,2,0}$ & 
  $C_{2,2,2}^{-1,-1,2}$ & 
   $C_{2,2,2}^{1,1,-2}$ & 
   $C_{2,2,2}^{-1,1,0}$ &
    $C_{2,2,2}^{-1,2,-1}$
  \\ [0.5ex] 
 \hline\hline
  $   \frac{1}{7} \sqrt{\frac{5}{\pi}}$ & 
 $ -\frac{1}{7} \sqrt{\frac{5}{4\pi}}  $& 
 $\frac{1}{7} \sqrt{\frac{15}{2\pi}} $ & 
 $\frac{1}{7} \sqrt{\frac{15}{2\pi}}$ & 
 $-\frac{1}{4} \sqrt{\frac{5}{\pi}}$ &
 $   \frac{1}{14} \sqrt{\frac{30}{\pi}}$ 
 \\  [1ex] 
 \hline
\end{tabular}
\end{center}
 \begin{center}
 \begin{tabular}{|| c | c | c  | c | c | c | c | c | c ||} 
 \hline
  $C_{2,2,4}^{0,0,0}$ & 
  $C_{2,2,4}^{2,2,-4}$  & 
  $C_{2,2,4}^{-2,-2,4}$ & 
  $C_{2,2,4}^{-2,2,0} $ & 
  $C_{2,2,4}^{-1,-1,2} $ & 
  $C_{2,2,4}^{1,1,-2}$ & 
  $C_{2,2,4}^{-1,1,0}$ &
  $C_{4,2,2}^{-1,2,-1}$ & 
   $C_{4,2,2}^{3,-2,-1}$   
\\ [0.5ex] 
 \hline\hline
 $  \frac{3}{7} \sqrt{\frac{1}{\pi}} $ & 
 $ 5 \sqrt{\frac{1}{70 \pi}} $ & 
 $5  \sqrt{\frac{1}{70\pi}}  $& 
 $ \frac{1}{7}  \sqrt{\frac{1}{2\pi}}  $  &
 $\frac{1}{7} \sqrt{\frac{10}{ \pi}} $ & 
 $\frac{1}{7} \sqrt{\frac{10}{ \pi}}$ &
 $\frac{2}{7} \sqrt{\frac{1}{ \pi}}$ &
 $ - \frac{5}{14} \sqrt{\frac{1}{5\pi}} $&
 $ -\frac{5}{2} \sqrt{\frac{1}{35\pi}}  $
 \\  [1ex] 
 \hline
\end{tabular}
\end{center}

\section{Calculation of the matrices $\gamma_{\mu}$}\label{gamma}
%In 3D case without crystal field $f_{\mu} ({\bf k})= Y_2^m(\hat{k})$. 
Following a similar procedure as the one followed to calculate $F$, we have for plane waves
\begin{eqnarray} \label{gam3}
\gamma_{\mu}({\bf r},\omega) &=&\sum_{\bf k} \sum_{l^{\prime},m^{\prime}} (i)^{l^{\prime}} j_{l^{\prime}}(kr) Y_{l^{\prime}}^{m^{\prime}}(\hat{r}) Y_{l^{\prime}}^{m^{\prime}}(\hat{k}) f_{\mu}(\hat{k})^* G^0({\bf k}, \omega) 
 \simeq  - 4 \pi \nu \int d \xi j_2 \left(\frac{\xi}{{\rm v}_F} + k_F r\right)  G^0( \xi, \omega)f_{\mu}(\hat{r})
 \\
&= &  4 \pi \nu   \left[ \left( \omega \tau_0 + \Delta \tau_x \right) I^{(2)}_1(r,\omega) + \tau_z I^{(2)}_2(r,\omega) \right] f_{\mu}(\hat{r}).
\end{eqnarray}
 Notice that the asymptotic behavior is dominated by the terms $\propto 1/(k_F r)$ in the 3D case and as $1/\sqrt{k_F r}$ in the 2D one.

\section{Evaluation of integrals $I^{(l)}_1(r,\omega)$ and $I^{(l)}_2(r,\omega)$}\label{integrals}
%\subsection{3D}
We present the evaluation of integrals defined in Eq. (\ref{inte}) for the cases $l=0,2,4$.  The corresponding Bessel functions are
\begin{equation}
j_0(u)=\frac{\sin u}{u},\;\;\;\;\;\;\;\;\;j_2(u)= \frac{\sin u}{u}\left[ \frac{3}{u^2}- 1 \right] - \frac{3 \cos u}{u^2},\;\;\;\;\;\;\;\;\; j_4(u)=\frac{\sin u}{u}\left[ \frac{105}{u^4}-\frac{45}{u^2}+ 1 \right]+\cos u \left[ \frac{10}{u^2}-\frac{105}{u^4} \right],
\end{equation}
with $u= \left( \xi/v_F + k_F \right) r$.

\subsubsection{l=0}
The integrals we have to evaluate  can be expressed as follows
\begin{equation}
I^{(0)}_{j}= \frac{r}{v_F} \int du \frac{ \sin u \left[ v_F (u/r -k_F) \right]^{j-1} }{(u-u_+)(u-u_-)u},
\end{equation}
with 
\begin{equation}\label{upm}
u_{\pm} =k_F r \pm i r \sqrt{\Delta^2 - \omega^2} /v_F.
\end{equation}
 We proceed as follows
\begin{equation}
I^{(0)}_{1} = \frac{r}{v_F} \frac{1}{2i} \left[ \int_{C_1} du \frac{e^{i u}}{(u+i\eta) (u-u_+) (u-u_-)} - \int_{C_2}  du\frac{e^{-i u}}{(u-i\eta) (u-u_+) (u-u_-)} \right],
\end{equation}
where $C_1 \; (C_2)$ are complex contours that runs along the real axis and close in semicircles  with radius $R \rightarrow \infty$ in the upper (lower) semiplane. The result from Cauchy theorem is
\begin{equation}
I^{(0)}_{1} = \frac{\pi r}{v_F}  \left[ \frac{e^{i u_+}}{(u_++i\eta)  (u_+-u_-)} + \frac{e^{-i u_-}}{(u_--i\eta) (u_--u_+) } \right]=\frac{\pi}{\sqrt{\Delta^2 -\omega^2}} 
\frac{e^{- \frac{r}{v_F}\sqrt{\Delta^2-\omega^2}}}{(k_F r)^2+(\Delta^2-\omega^2)(r/v_F)^2} \left( k_F r \sin k_F r - \frac{r}{v_F} \sqrt{\Delta^2-\omega^2} \right).
\end{equation}
We proceed similarly with the other integral. The result is
\begin{equation}
I^{(0)}_j \simeq \lambda_j(r,\omega) \kappa^{(0)}_j(k_F r), \;\;\;\;\;\;\;\ r < \xi_0,
\end{equation}
where $\xi_0= v_F/\Delta$ is the coherence length of the superconductor. 
We have defined
\begin{equation} \label{kappa0}
\kappa^{(0)}_{1} (k_F r) =  \frac{\sin k_F r}{k_F r}, \;\;\;\;\;\;\;\kappa^{(0)}_{2} (k_F r)  =  \frac{\cos k_F r}{(k_F r)}.
\end{equation}
and 
\begin{equation} \label{lambda12}
\lambda_1(r,\omega)= \frac{\pi}{\sqrt{ \Delta^2 - \omega^2}} e^{-\sqrt{1 - \left(\frac{\omega}{\Delta}\right)^2} \frac{ r}{\xi_0}}, \;\;\;\;\;\;\;\;\; \lambda_2(r,\omega)= \pi e^{-\sqrt{1 - \left(\frac{\omega}{\Delta}\right)^2} \frac{ r}{\xi_0}}.
\end{equation}

\subsubsection{l=2}
We express the integrals  for $j=1,2$ as $I^{(2)}_1(r,\omega)=I^{(2)}_{1,1}+I^{(2)}_{1,2}+I^{(2)}_{1,3}$, and $I^{(2)}_2(\omega)=I^{(2)}_{2,1}+I^{(2)}_{2,2}+I^{(2)}_{2,3}$ being
\begin{eqnarray}
I^{(2)}_{j,1}&=& \frac{r}{v_F} \mbox{PV} \int du \frac{3 \sin u \left[ v_F (u/r -k_F) \right]^{j-1} }{(u-u_+)(u-u_-)u^3},\nonumber \\
I^{(2)}_{j,2}&=&- \frac{r}{v_F} \int du \frac{ \sin u \left[ v_F (u/r -k_F) \right]^{j-1} }{(u-u_+)(u-u_-)u},\nonumber \\
I^{(2)}_{j,3}&=&- \frac{r}{v_F} \mbox{PV} \int du \frac{3 \cos u \left[ v_F (u/r -k_F) \right]^{j-1} }{(u-u_+)(u-u_-)u^2},
\end{eqnarray}
with $u_{\pm} $ defined in Eq. (\ref{upm}). 
We proceed as in the case with $l=0$.
 The results are
 \begin{equation}
I^{(2)}_{j} \simeq \lambda_j(r,\omega) \kappa^{(2)}_{j}(k_F r),\;\;\;\;\;\;\;\;\;\;\;\;\;\;\;\;\;\;\;\; \kappa^{(2)}_{j}(k_F r)= \sum_{i=1}^3 \kappa^{(2)}_{j,i}(k_F r)
\end{equation}
with 
\begin{eqnarray} \label{kappa2}
\kappa^{(2)}_{1,1}(k_F r)& = &3   \frac{\sin k_F r}{(k_F r)^3}, \;\;\;\;\;\;\;
\kappa^{(2)}_{1,2}(k_F r) = - \frac{\sin k_F r}{k_F r}, \;\;\;\;\;\;\;
%\nonumber \\
\kappa^{(2)}_{1,3}(k_F r)  =  -  3  \frac{\cos k_F r}{(k_F r)^2}, \nonumber \\
%\end{eqnarray}
%For the terms of the integral $I_2^{(2)}$ we get
%\begin{eqnarray}
\kappa^{(2)}_{2,1}(k_F r) & = & 3  \frac{\cos k_F r}{(k_F r)^3}, \;\;\;\;\;\;\;\;\;
%\nonumber \\
\kappa^{(2)}_{2,2}(k_F r)  =  -  \frac{\cos k_F r}{(k_F r)}, \;\;\;\;\;\;\;\;\;
%\nonumber \\
\kappa^{(2)}_{2,3}(k_F r)  = 3  \frac{\sin k_F r}{(k_F r)^2}.
\end{eqnarray}
The functions $\lambda_{1,2}$ are defined in Eq. (\ref{lambda12}).

\subsubsection{l=4}
We proceed as in the previous case and split the integrals as $I^{(4)}_j(r,\omega)=\sum_{i=1}^5 I^{(4)}_{j,i}, \; j=1,2 $, being
\begin{eqnarray}
I^{(4)}_{j,1}&=& 105 \frac{r}{v_F} \mbox{PV} \int du \frac{ \left[ v_F (u/r -k_F) \right]^{j-1} \sin u}{(u-u_+)(u-u_-)u^5},\nonumber \\
I^{(4)}_{j,2}&=&- 45 \frac{r}{v_F}\mbox{PV} \int du \frac{ \sin u \left[ v_F (u/r -k_F) \right]^{j-1} }{(u-u_+)(u-u_-)u^3},\nonumber \\
I^{(4)}_{j,3}&=& \frac{r}{v_F} \mbox{PV} \int du \frac{ \sin u \left[ v_F (u/r -k_F) \right]^{j-1} }{(u-u_+)(u-u_-)u},\nonumber \\
I^{(4)}_{j,4}&=&10  \frac{r}{v_F} \mbox{PV} \int du \frac{\cos u \left[ v_F (u/r -k_F) \right]^{j-1} }{(u-u_+)(u-u_-)u^2},\nonumber \\
I^{(4)}_{j,5}&=& -105 \frac{r}{v_F} \mbox{PV} \int du \frac{ \cos u \left[ v_F (u/r -k_F) \right]^{j-1} }{(u-u_+)(u-u_-)u^4}.
\end{eqnarray}
The results are
 \begin{equation}
I^{(4)}_{j} = \lambda_j(r,\omega) \kappa^{(4)}_{j}(k_F r),\;\;\;\;\;\;\;\;\;\;\;\;\;\;\;\;\;\;\;\; \kappa^{(4)}_{j}(k_F r)= \sum_{i=1}^5 \kappa^{(4)}_{j,i}(k_F r),
\end{equation}
with  
\begin{eqnarray} \label{kappa4}
\kappa^{(4)}_{1,1}(k_F r)& = &105  \frac{\sin k_F r}{(k_F r)^5}, \;\;\;
  \kappa^{(4)}_{1,2}(k_F r) = - 45 \frac{\sin k_F r}{(k_F r)^3}, \;\;\;
\kappa^{(4)}_{1,3}(k_F r)  =    \frac{\sin k_F r}{(k_F r)}, \;\;\;
\kappa^{(4)}_{1,4}(k_F r)  =   10 \frac{\cos k_F r}{(k_F r)^2}, \;\;\;
\kappa^{(4)}_{1,5}(k_F r)  =  - 105  \frac{\cos k_F r}{(k_F r)^4}, \nonumber \\
\kappa^{(4)}_{2,1}(k_F r) & = &105  \frac{\cos k_F r}{(k_F r)^5}, \;\;\;
 \kappa^{(4)}_{2,2}(k_F r) = - 45  \frac{\cos k_F r}{(k_F r)^3}, \;\;\;
\kappa^{(4)}_{2,3}(k_F r)  =     \frac{\cos k_F r}{(k_F r)}, \;\;\;
\kappa^{(4)}_{2,4}(k_F r)  =    -10 \frac{\sin k_F r}{(k_F r)^2}, \;\;\;
\kappa^{(4)}_{2,5}(k_F r)  =   105  \frac{\sin k_F r}{(k_F r)^4}, \nonumber
\end{eqnarray}

\end{widetext}
%%%%%%%%%%%%%%%%%%%%%%%%%%%%%%%%%%%%%%%%%%%%%%%%%%%%%%%%%%%%%%%%%

%%%%%%%%%%%%%%%%%%%%%%%%%%%%%%%%%%%%%%%%%%%%%%%%%%%%%%%%%%%%%%%%%

\end{document}